\documentclass[prc,preprint,superscriptaddress,nofootinbib]{revtex4}
\usepackage{amsfonts}
\usepackage{amssymb}
\usepackage{amsfonts}
\usepackage{amsmath}
\usepackage{amssymb}
\usepackage{amsthm}
\usepackage{epsf}
\usepackage[dvips]{graphicx}

\def\bp{{\mbox{\boldmath$p$}}}

\def\bphi{\mbox{\boldmath{$\phi$}}}
\begin{document}

\title{Analytical properties of the quark propagator from truncated Dyson-Schwinger equation in complex Euclidean space}

\author {S.~M. Dorkin}
\affiliation{Bogoliubov Lab.~Theor.~Phys., 141980, JINR, Dubna,
 Russia}
\affiliation{International University Dubna, Dubna, Russia }

 \author{L.~P. Kaptari}
 \affiliation{Bogoliubov Lab. Theor. Phys., 141980, JINR, Dubna,
 Russia}
\affiliation{Helmholtz-Zentrum Dresden-Rossendorf, PF 510119, 01314
Dresden, Germany}
\author {T.~Hilger}
\affiliation{Helmholtz-Zentrum Dresden-Rossendorf, PF 510119, 01314
Dresden, Germany}
\affiliation{Institut f\"ur Theoretische Physik, TU Dresden, 01062 Dresden, Germany}

\author { B.~K\"ampfer}
\affiliation{Helmholtz-Zentrum Dresden-Rossendorf, PF 510119, 01314
Dresden, Germany}
\affiliation{Institut f\"ur Theoretische Physik, TU Dresden, 01062 Dresden, Germany}

\begin{abstract}

  In view  of the mass spectrum
  of heavy mesons in vacuum  the
  analytical properties of the solutions of the truncated Dyson-Schwinger equation
  for the quark propagator
  within the rainbow approximation are analysed in some detail. In
  Euclidean space,  the quark propagator is not an analytical function
  possessing,  in general,  an infinite number of singularities (poles)
   which   hamper to solve the Bethe-Salpeter equation.
  However, for light mesons (with masses $M_{q\bar q} \le 1$ GeV) all singularities
  are located outside the region within which the Bethe-Salpeter equation is defined.
   With an increase of the considered meson masses this region enlarges and already at
   masses $\ge $ 1 GeV, the poles of propagators of $u,d$ and $s$
   quarks  fall within the integration domain of the
   Bethe-Salpeter equation. Nevertheless,
   it is established that for meson masses up to $M_{q\bar q} \simeq 3$ GeV
   only the first, mutually complex conjugated,  poles contribute to the solution.
  We argue that, by knowing the position of the poles and their
   residues,   a reliable parametrisation of the quark propagators can be
   found and used in numerical procedures of solving the Bethe-Salpeter equation.
   Our analysis is  directly related to the future physics programme at FAIR
   with respect to open charm degrees of freedom.

\end{abstract}

\maketitle
\section{Introduction}

 The description of mesons as quark-antiquark bound states within the framework of the Bethe-Salpeter (BS) equation
 with momentum dependent quark mass functions, determined by the Dyson-Schwinger (DS) equation, is able
 to explain successfully many spectroscopic data, such as meson masses
 \cite{rob-1,Maris:1999nt,Maris:2003vk,Holl:2004fr,Blank:2011ha},   electromagnetic properties
 of pseudoscalar mesons and their radial excitations~\cite{Jarecke:2002xd,Krassnigg:2004if,Roberts:1994hh,Roberts:2007jh})
 and  other   observables~\cite{Maris:2000sk,Maris:1999bh,ourFB,wilson,JM-1,JM-2,rob-2,tandy1,David,Alkofer,fisher,Roberts}.
 Contrarily to purely  phenomenological models, like  the quark bag  model,
 the presented formalism  maintains   important features of QCD, such as dynamical chiral
 symmetry breaking, dynamical quark dressing, requirements of the renormalization group theory etc., cf.~Ref.~\cite{physRep}.
 The main ingredients here are  the full quark-gluon vertex function and  the dressed gluon
 propagator,  which are entirely determined  by  the running coupling
 and the bare quark mass parameters. In principle, if one were able to solve the Dyson-Schwinger equation,
 the approach would  not depend on  any additional parameters. However, due to  known technical problems,
 one restricts oneself to  calculations of the few first terms of the perturbative series,
 usually up to  the one-loop approximation. The obtained results, which formally  obey  all the
 fundamental requirements of the theory, are then considered as exact ones,  however, with effective parameters.
 This is known as the rainbow-ladder approximation of the DS equation.
 The merit of the approach is that, once the effective parameters are fixed,
 the whole spectrum of known mesons is supposed to  be described   on the same footing,
 including also excited states.

It should be noted that there exists other approaches based on the same physical ideas
but not so sophisticated, e.g. employing simpler interactions, such as a  separable interaction for the effective
coupling \cite{David}. Such approaches  describe also fairly well  properties of light mesons,
nevertheless, the investigation of  heavier mesons and excited states,
consisting even of light $u$, $d$ and $s$  quarks, requires implementations of
more accurate numerical methods to solve the corresponding equations.

In the present paper we investigate  the prerequisites to the interaction
kernel of the combined Dyson-Schwinger and Bethe-Salpeter  formalisms to describe
 the meson mass spectrum including heavier mesons and excited states as well.
Particular attention is paid to the charm sector which, together with the
 baryon spectroscopy, is a major topic in the FAIR research programme.
Two large collaborations at  FAIR \cite{CBM,PANDA} plan precision measurements. Note that  it becomes now
possible to experimentally investigate  not only the mass spectrum of the mentioned mesons, but also
different processes of their decay, which are directly connected with fundamental QCD problems
(e.g., $U(1)$ axial anomaly, transition form factors etc.) and with the challenging  problem of
 changes of meson characteristics
at finite temperatures and densities. The latter is  crucial in understanding the di-lepton yields in nucleus-nucleus collisions
at, e.g., HADES.  All these circumstances  require  an adequate theoretical foundation to describe
the meson spectrum and the meson covariant wave functions (i.e. the BS partial amplitudes) needed in   calculations
of the relevant  Feynman diagrams and observables.

   Due to   the
  momentum dependence of the quark mass functions,  the BS equation requires
  an analytical continuation of the quark propagators in the
  complex plane of Euclidean momenta which can be achieved either by corresponding numerical
  continuations of the solution obtained along the real axis or by solving directly
  the DS equation in the complex domain of validity
  of the equation itself.  With increasing  meson mass $M_{q\bar q}$ this region enlarges,  and already at
   masses $M_{q\bar q}\ge~1$ GeV, singularities of light-quark propagators fall within the
  domain the BS equation to be solved.
   It is found that even so, only the few first poles contribute to the solution.
   To analyse the analytical properties of the quark propagators we solve the
   DS equation in the rainbow ladder approximation by making use of the hyperspherical
   harmonics basis to decompose the propagators and the corresponding potential
   and solve numerically the resulting DS equations for the coefficients of such a
   decomposition. The further  analysis of the   solution
   is based on a combined application of  Rouch\'e's theorem and a graphical
   representation of the inverse propagators  as  vortex fields of the corresponding
   complex functions. Since the main goal of our analysis is the use of the quark propagator
   functions evaluated at such complex
   momenta for which the are needed in the BS equation, we focus our attention on this region of Euclidean space.

 Our paper is organized as follows. In Sec.~\ref{s:bse}, Subsecs.~\ref{Bet}~and~\ref{Dys}  we briefly discuss the truncated
 BS and DS equations relevant to describe the mesons as quark-antiquark
 bound states. The rainbow approximation for the DS equation kernel is  introduced
 in Subsec.~\ref{choos}, and  the domain of the complex plane of
 Euclidean momenta, where the solutions are sought, is specified in Subsec.~\ref{domain}.
 The explicit form of the DS equation to be solved
 in the rainbow ladder approximation with effective infrared kernel is formulated
  and the corresponding
 numerical solution for space-like momentum is discussed in~Subsec.~\ref{explic}.
 Section~\ref{sec3}
 is entirely dedicated to the solution of the truncated DS equation for complex  momenta.
 In Subsec.~\ref{subsecphi} we consider the analytical continuation of the solution for
 complex momenta along rays with
constant angle. It is found that in the right hemisphere of complex momenta squared, $p^2$,
 the solutions of DS equation are analytical functions so that all singularities of the quark propagator
are to be searched for in the left hemisphere. The subsequent Sections are aimed at investigations of
the solutions in the time-like domain of $p^2$ determined by the BS equation for mesons with masses
$M_{q\bar q}\lesssim 2-3$ GeV:
 In Subsec.~\ref{subsectimelike} we present the solutions and the propagator functions for the $c$
 quark, for which it was found that they are analytical functions in the considered region. A reliable parametrization
 is found, allowing to facilitate the numerical procedure of solving the BS equation
 for charmed mesons. A thorough investigation of the pole structure of propagators of light quarks is presented in
 Sec.~\ref{polestruc}. By combining  Cauchy and Rouch\'e's theorems with an analysis of the force lines of vector fields
 of propagators as complex functions, the position of first few poles and the corresponding residues  the
 propagator functions  are found with a good accuracy. The dependence of pole locations in the complex
 momentum plane  on the bare quark mass is analysed as well.
The impact of the ulraviolet term in the interaction kernel is briefly discussed in Sec.~\ref{UVimpact}. Summary and conclusions
are collected in Sec.~\ref{summary}.

\section{Bethe-Salpeter and Dyson-Schwinger Equations}
\label{s:bse}
\subsection{Bethe-Salpeter equation}
\label{Bet}
To determine the bound-state mass of a quark-antiquark pair one needs to solve the Bethe-Salpeter equation, which
in the ladder approximation  (hereafter referred to as truncated Bethe-Salpeter (tBS) equation)  and
in Euclidean space  reads
\begin{eqnarray}
    \Gamma(P,p) =   -\frac 43  \int \frac {d^4k}{(2\pi)^4}
    \gamma_{\mu}S(k_1) \Gamma(P,k) S(k_2)\gamma_{\nu}
    \left [g^2    {\cal D}_{\mu \nu}(p-k)\right ] \: ,
\label{bse}
\end{eqnarray}
with $\Gamma$ being the BS vertex function and
$S^{-1}(k_{1})=\left (i\gamma\cdot k_1 +m_1\right)$  and $S^{-1}(k_{2})=\left (i\gamma\cdot k_2 +m_2\right)$ are the
inverse  propagators of two quarks, which interact via gluon exchange encoded in
$\left [g^2    {\cal D}_{\mu \nu}(p-k)\right ]$.
The vertex
function $\Gamma (P,k)$ is a $4\times 4$ matrix and, therefore, may contain
16 different functions.
The general structure of the  vertex functions
describing bound states of spinor particles  has been
investigated in detail, for example, in \cite{ourUmn,kubis,dorByer}. To release from the matrix structure, the vertex
function  $\Gamma$ is  expanded into functions which in turn are determined by angular momentum and parity of the corresponding
meson known as the spin-angular harmonics~\cite{kubis,dorByer,ourFB,ourFB1}:
\begin{eqnarray}
    \Gamma(p_4,\bp)&=&\sum\limits_\alpha g_\alpha(p_4,|\bp|) \,{\cal
    T}_\alpha(\bp) \label{spex} \: .
\end{eqnarray}
With Eq.~(\ref{spex}) it can be shown~\cite{dorByer,ourFB1} that the integral matrix form of the  BS equation
(\ref{bse}) can be reduced to a system of  ordinary algebraic equations.
 It should be noted that, if one  would consider the meson as a bound state of two quarks with constant masses,
 i.e. if the meson is treated within the Bethe-Salpeter formalism with effective quark masses, then
 the scalar part of the product of quark propagators becomes purely real and free of any singularities.
 In such a case the quark masses appear as effective parameters, which need to be
 different for different mesons. In a more realistic case, where mesons are described on
 a common footing as bound states of dynamically
 dressed quarks, the corresponding "masses" are represented by  rather complicate functions of the momenta, and
 the product of two propagators in the above  tBS equation remains complex, even in Euclidean space.
 That means  prior proceeding in solving the tBS equation, one needs to know  the
 analytical properties of quark propagators in the complex Euclidean momentum space.
\subsection{Dyson-Schwinger equation}

\label{Dys}

The coupled equations of the quark propagator $S$, the gluon propagator $D_{\mu\nu}$ and the vertex
function $\Gamma_\mu$ are often considered as integral formulation being equivalent to full QCD. While
there are attempts to solve this coupled set of DS equations by some numerical procedures, for certain purposes
some approximations~\cite{fisher,Maris:2003vk,physRep}  are appropriate. Being interested in dealing with mesons as
quark-antiquark bound states within utilization of the tBS equation, one has to provide the quark propagator
which depends on the gluon propagator and vertex as well, which in turn depend on the quark propagator.
Leaving  a detailed discussion of the variety of approaches in dressing
 of the gluon propagator and vertex function in DS equations
 (see e.g.~Refs.~\cite{Fischer:2008uz,PenningtonUV} and references therein
 quoted) we mention only that in solving the DS equation for the quark propagator one usually employs truncations of the
 exact interactions and replaces the gluon propagator combined with the vertex
 function by effective interaction kernels. This leads to the truncated Dyson-Schwinger (tDS) equation for
 the quark propagator.
In concrete calculations the choice of the form of the effective interaction is inspired by
results from calculations of Feynman diagrams within  pQCD maintaining
requirements of symmetry and asymptotic behaviour  already implemented, cf.~Refs.~\cite{Maris:2003vk,Roberts,physRep,PenningtonUV}.
 The  results of such calculations, even in the simplest case of accounting only for one-loop
 diagrams with proper regularization
 and renormalization procedures,  are rather cumbersome for further use in
 numerical calculations, e.g. in the framework of BS or Fadeev equations.
 Consequently, in practice,  the wanted exact results are
 replaced by  parametrizations of the vertex and gluon propagator. Often one
 employs  an  approximation which corresponds to one-loop calculations of diagrams with
 the  full  vertex function $\Gamma_\nu$,
 substituted by the free one,
  $\Gamma_\nu(p,k)\rightarrow\gamma_\nu$ (we suppress the color structure and
  account cumulatively for the strong coupling later on).
  In  Euclidean space the quark propagator obeys then  the   tDS equation
\begin{eqnarray}
S^{-1}(p)= S_0^{-1}(p) + \frac 43 \int \frac
{d^4 k }{(2\pi)^4} \left[g^2 {\cal D}_{\mu \nu}(p-k) \right]\gamma_{\mu} S(k)
\gamma_{\nu}\: ,
\label{sde}
\end{eqnarray}
where $S_0^{-1}=i \gamma \cdot p + m_q$ and ${m}_q$ is the bare current  quark mass. To emphasize
the replacement of combined gluon propagator and vertex we use, as in Eq.~(\ref{bse}), the notation
$[g^2D_{\mu\nu}]$, where an additional power of $g$ from the second undressed vertex is
included.  For a consistent
treatment  of  dressed quarks and their bound states, the dressed gluon propagator
$[g^2 {\cal D}_{\mu \nu}(p-k)]$ must
be   the same in the tBS, (\ref{bse}),  and the tDS equations, (\ref{sde}).

\subsection{Choosing an interaction kernel}\label{choos}
  This truncation, known also as the
  ladder rainbow approximation, has been   widely used  to study the physics of dynamical chiral symmetry breaking
 \cite{Maris:1999nt}, decay constants \cite{Maris:2003vk,Holl:2004fr,Jarecke:2002xd,Blank:2011ha,Krassnigg:2004if} and other   observables
 \cite{Roberts:1994hh,Maris:2000sk,Maris:1999bh,ourFB} and has been found to
 provide a good agreement with experimental data.
 The employed vertex-gluon kernel in the rainbow approximation
 is also known as the Maris-Tandy (MT) model \cite{Maris:1999nt}.
 It  is chosen here in the form  \cite{Maris:1999nt,Maris:2003vk,Alkofer,Roberts}
\begin{eqnarray}
 g^2(k^2) {\cal D}_{\mu \nu} (k^2) =
    \left(
        \frac{4\pi^2 D k^2}{\omega^2} e^{-k^2/\omega^2}
        + \frac {8\pi^2 \gamma_m F(k^2)}{
            \ln[\tau+(1+\frac{k^2}{\Lambda_{QCD}^2})^2]}
    \right)
    \left( \delta_{\mu\nu}-\frac {k_{\mu} k_{\nu}}{k^2} \right) \, ,
\label{phenvf}
\end{eqnarray}
 where the first term originates from  the effective infrared (IR) part of the interaction
 determined by soft non-pertubative effects, while the second one ensures the correct
 ultraviolet (UV) asymptotic behaviour of the QCD running constant.  A detailed investigation of
 the interplay of these two terms
 has shown~\cite{souglasInfrared,Blank:2011ha} that the IR part  is dominant for light $u$,  $d$ and $s$ quarks
 with a decreasing role for heavier quark  masses ($c$ and $b$) for which the
UV part may be quite important in forming  meson masses $M_{q\bar q}> 3-4$ GeV as bound states.
  In the present paper we focus on mesons with
  masses $M_{q\bar q}< (2-3)$ GeV which consist of at least one light quark and for which the
  UV term can be seemingly neglected. However, in spite of its minor contribution
  to meson masses, formally the UV term guaranties
  the correct asymptotic behaviour of the kernel and in principle it must be included
  even in case of light quarks  forming mesons.
\subsection{The relevant region for the \lowercase{t}DS equations}\label{domain}
\begin{figure}[hbt]             
\includegraphics[scale=0.6 ,angle=0]{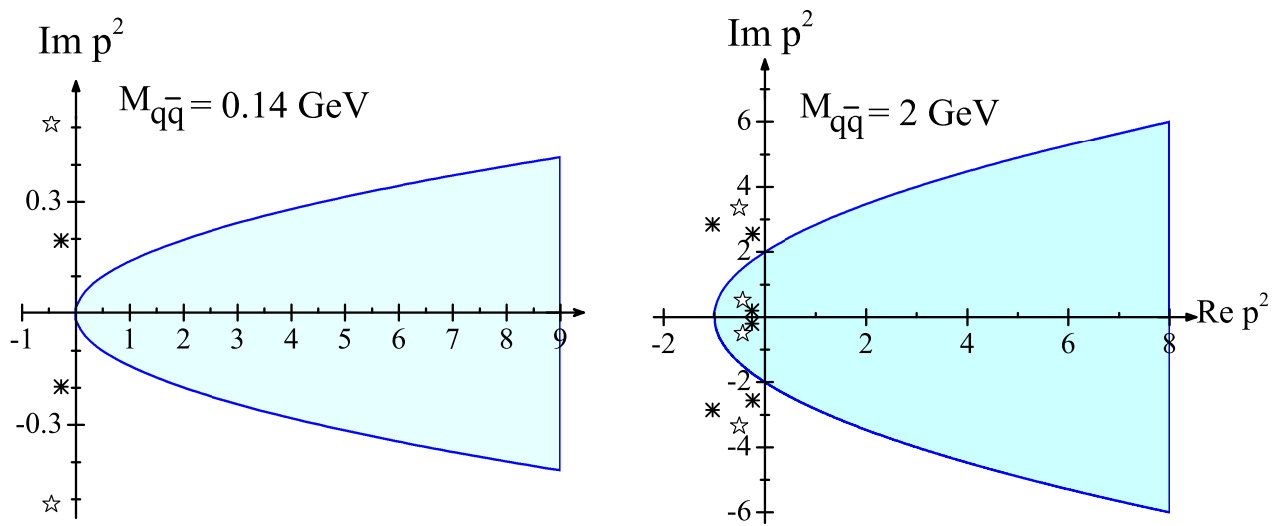}
\caption{(Color online) The  Euclidean space, where  numerical solutions of  the tBS
 and  tDS equations are sought. The quark inverse propagator part, $ \Pi(p) =p^2 A^2(p)+B^2(p))$, where
 $p^2=-M^2_{q\bar q}/4 + k^2 \pm iM_{q\bar q} k \cos \chi$, entering  the tBS
 equation is defined within the colored areas of the corresponding parabolas ($-1\le\cos\chi\le 1$). Left
 (right) panel:  the integration domain for  $M_{q\bar q}=140\, MeV$ ($M_{q\bar q}=2$ GeV).
 The integration domain is restricted to $k_{max}\leq 3$ GeV/c; for larger  values of $k$
 all the partial Bethe-Salpeter amplitudes  are already negligible small (see, e.g. Ref.~\cite{dorByer}).
 For  light mesons (left panel, $m_q=0.005$ GeV) there are no singularities in the colored tBS
 integration domain. In the right panel  we display the   first six self-conjugated poles of
propagators of light, $u$,  $d$ quarks (asterisks) and the first four poles for the $s$ quarks (open stars)
 which enter the solution of the tBS equation. The pole positions are quoted in Table~\ref{tb3}. Axes are in
 units of ${\rm (GeV/c)}^2$.}
\label{parabola}
\end{figure}

  In a first step in our investigation  we disregard the UV term   and study the analytical properties of the
  corresponding propagator functions obtained with the IR term solely (cf.~also Refs.~\cite{Blank:2011ha,Roberts:2007ji,Alkofer}).
  Then we reconcile the results with the UV term in the kernel in Section~\ref{UVimpact}. As mentioned above, we are interested
  in the analytical structure of the propagator function $S(p)$  inside and
  in the  neighbourhood of the complex momentum region in Euclidean space dictated by the tBS equation. This
  momentum region is displayed as the dependence of the imaginary part of the quark momentum squared Im\,$p^2$  on
  its real part Re\,$p^2$ determined by  the tBS equation.   In terms of the relative  momentum
  $k$ of the  two quarks residing in a meson the corresponding dependence
  is
  \begin{eqnarray}
  p^2 = -\displaystyle\frac{M_{q\bar q}^2}{4} + k^2 \pm i M_{q\bar q} k \label{parequation}
  \end{eqnarray}
   determining
  in the Euclidean complex momentum plane  a parabola
  Im\,$p^2=\pm~M_{q \bar q}\sqrt{{\rm Re}\,p^2~+~\frac{M_{q \bar q}^2}{4}}$ with  vertex at
  Im\,$p^2=0$ at Re\,$p^2 =-M_{q \bar q}^2/4 $  depending  on the
  meson mass $M_{q\bar q}$; the symmetry axis is the Re\,$p^2$
  axis, i.e. the parabola extends to Re\,$p^2\to\infty$. Two examples  are depicted in  Fig.~\ref{parabola}, for $M_{q \bar q}=0.14$ GeV (left) and
 $M_{q \bar q}=2$ GeV  (right).

  Note that regardless of the form of the interaction kernel, the
   investigation  of the analytical structure of the quark propagator
  is of  great importance, if the propagators exhibit singularities within the corresponding
  parabola, thus hampering the numerical procedure of solving the tBS equation.
  On the other side, the knowledge of the nature of singularities and their exact location
 in the complex plane will allow one to develop   effective  algorithms adequate for
 numerical calculations. For instance, if one determines exactly the
 domain of analyticity of the propagator functions, one can take advantage of the fact that
any  analytical function can always be approximated by rational
 complex  functions \cite{walsh}; then, one can   parametrize the
 integrand  in the tBS equation  by simple functions which allow to carry out some
 integrations analytically. Such parametrisations have been suggested in Ref.~\cite{souchlas} for
 meson masses $M_{q\bar q} <  1$ GeV  for which  the propagator functions  have been  found
  to be analytical.
Unfortunately, for larger meson masses the propagator functions
exhibit singularities within the domain of tBS integration and, as a consequence,
parametrizations by rational functions
are not possible.   Nevertheless, even  in this case, if the propagator  functions have only isolated poles
 with  known locations and  residues, then calculations can be significantly simplified by splitting the singular  functions
 into two terms, one analytical in the considered region the other one having
  a simple pole structure, as discussed below.
\subsection{Propagator functions}\label{explic}
 Coming back to the tDS equation~(\ref{sde}) we mention that it  is a four dimensional integral equation
 in matrix form. Simplifications can be achieved by exploiting
specific decompositions of the quark propagator.
  Then one decomposes the kernel over a complete set of basis functions,
 performs analytically  some angular integrations   and considers a new system of equations relative
to such a partial decomposition.
To be specific, we recall that the calculation of the renormalized Feynman diagrams leads to a
fermion propagator depending  on two  functions, e.g.
the renormalization constant $Z_2$ and the self-energy $\Sigma(p)$.
Instead of $Z_2$ and  $\Sigma(p)$ one can introduce other two quantities $A(p)$ and $B(p)$ or,
  alternatively, $\sigma_s(p)$ and $\sigma_v(p)$. The latter ones
   will be often addressed in the present paper to as the propagator
functions.
In terms of these functions the  dressed quark propagator $ S(p)$ reads  \cite{Roberts:2007ji,Maris:2003vk}
\begin{eqnarray}&&
    S^{-1}(p)= i \gamma \cdot p A(p)+ B(p),  \quad \quad S(p) =
-i\gamma \cdot p \sigma_v(p) +\sigma_s(p) \: ,\label{quarkpr1}
\end{eqnarray}
with
\begin{eqnarray}
\sigma_v(p)=\frac{A(p)}{p^2A(p)^2+B(p)^2}, \quad
\sigma_s(p)=\frac{B(p)}{p^2A(p)^2+B(p)^2}.
    \label{quarkpr}
\end{eqnarray}

The resulting system of equations to be solved  is
 (for details see Ref.~\cite{ourFB}; here we display only the formulae for the IR part of
 the kernel)
\begin{eqnarray}
    A(p)&=&1+2D \int   dk \frac { k^4}{p}
    \frac{A(k)}{k^2A^2(k)+B^2(k)}{\rm e}^{-\left(p-k\right)^2/\omega^2}
    \left\{ \frac{\left[ p^2 +k^2 +2\omega^2\right] }{kp} I_2^{(s)}(z) -2 I_1^{(s)}(z)
    \right\},\label{a}\\
     B(p)&=& m_q + 2D\int dk k^3   \frac{B(k)}{k^2A^2(k)+B^2(k)}
     {\rm e}^{-\left(p-k\right)^2/\omega^2}
    \left\{ \frac{\left[ p^2 +k^2\right] }{kp} I_1^{(s)}(z) -2 I_2^{(s)}(z)
    \right\},\label{ab1}
\end{eqnarray}
 where  $z=2pk/\omega^2$ and $I_n^{(s)}(z)$ are the  scaled
 (as emphasized by the label "(s)") modified Bessel
 functions  of the first kind defined as
 $I_{n}^{(s)}(z)\equiv \exp{(-z)}\ I_{n}(z),
 I_{n}^{(s)}(z\to\infty)= \displaystyle\frac{1}{\sqrt{2\pi z}}\left[1 -\displaystyle\frac{4n^2-1}{8z}\right ], 
\, I_{n}^{(s)}(z\to 0)=\left(\frac{z}{2}\right )^n{\rm e}^{-z} \Gamma(n+1)$.\\
The resulting  system  of one-dimensional integral equations Eqs.~(\ref{a}) and  (\ref{ab1}) can
be solved numerically  by an iteration method. Independent parameters are
$\omega$, $D$ and $m_q$. We find that the iteration
procedure   converges  rather fast and  practically does not depend  on
the choice of the trial start functions for $A(p)$ and $B(p)$.

\begin{figure}[!ht]
\includegraphics[scale=0.25 ,angle=0]{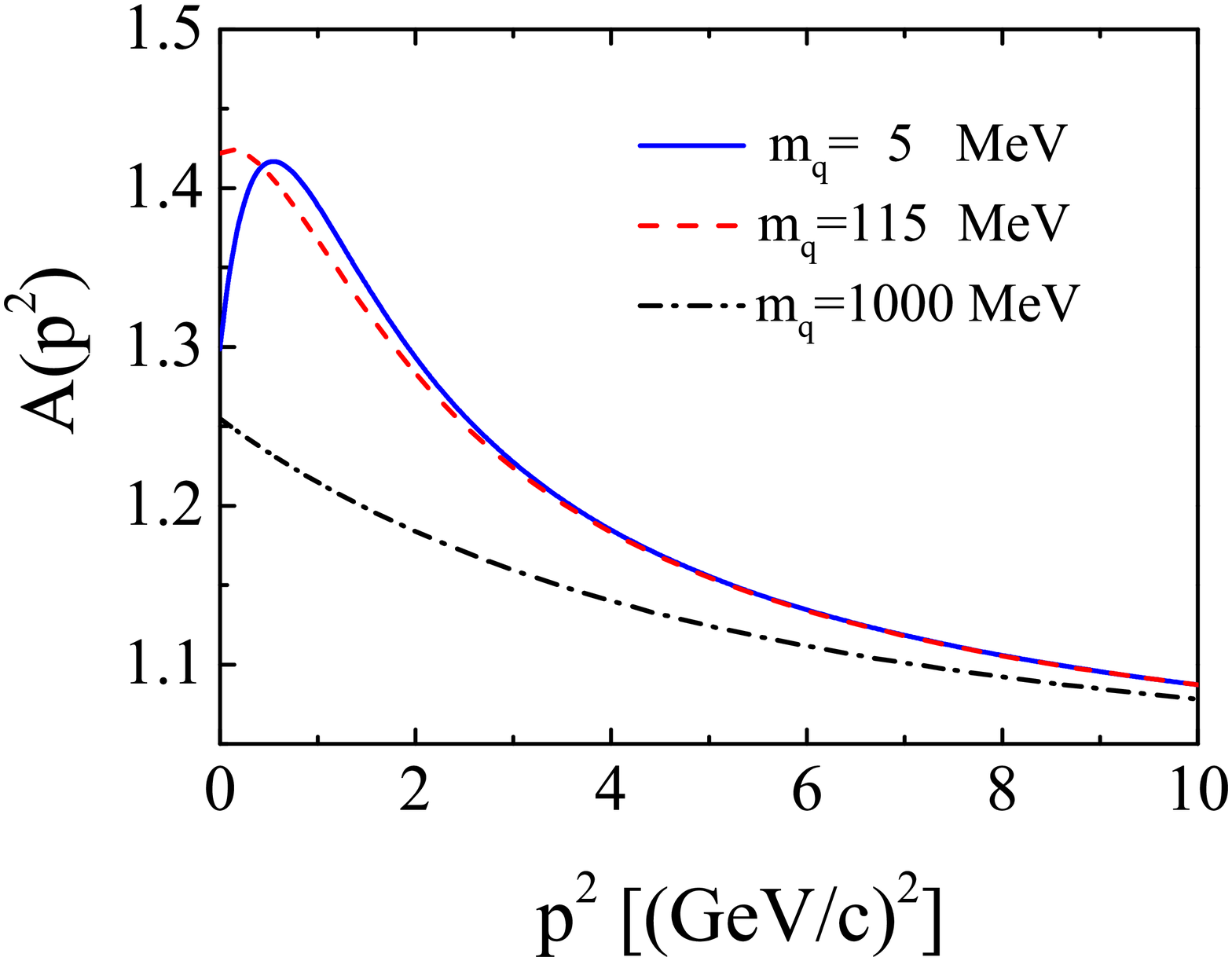}
\includegraphics[scale=0.25 ,angle=0]{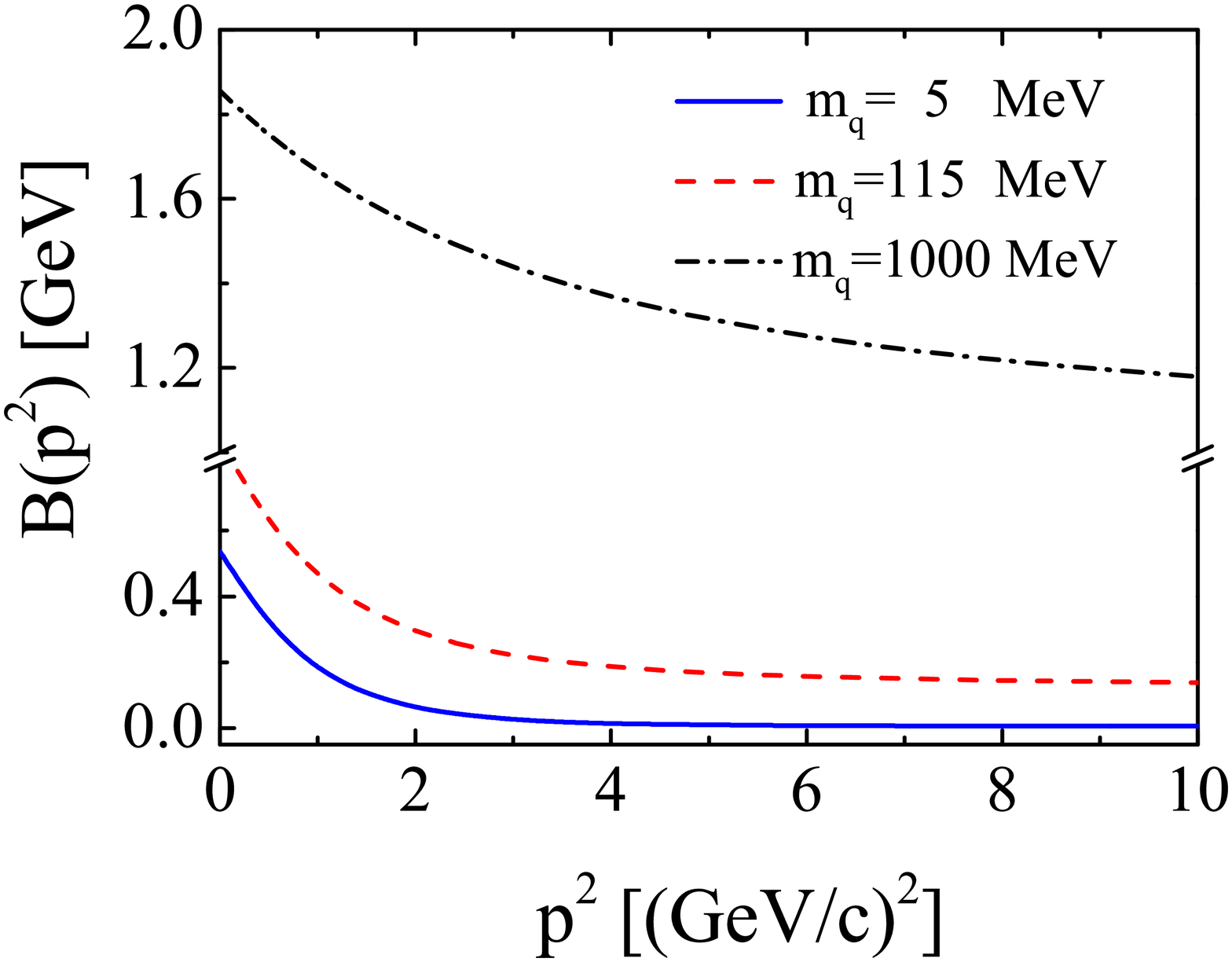}
\includegraphics[scale=0.25 ,angle=0]{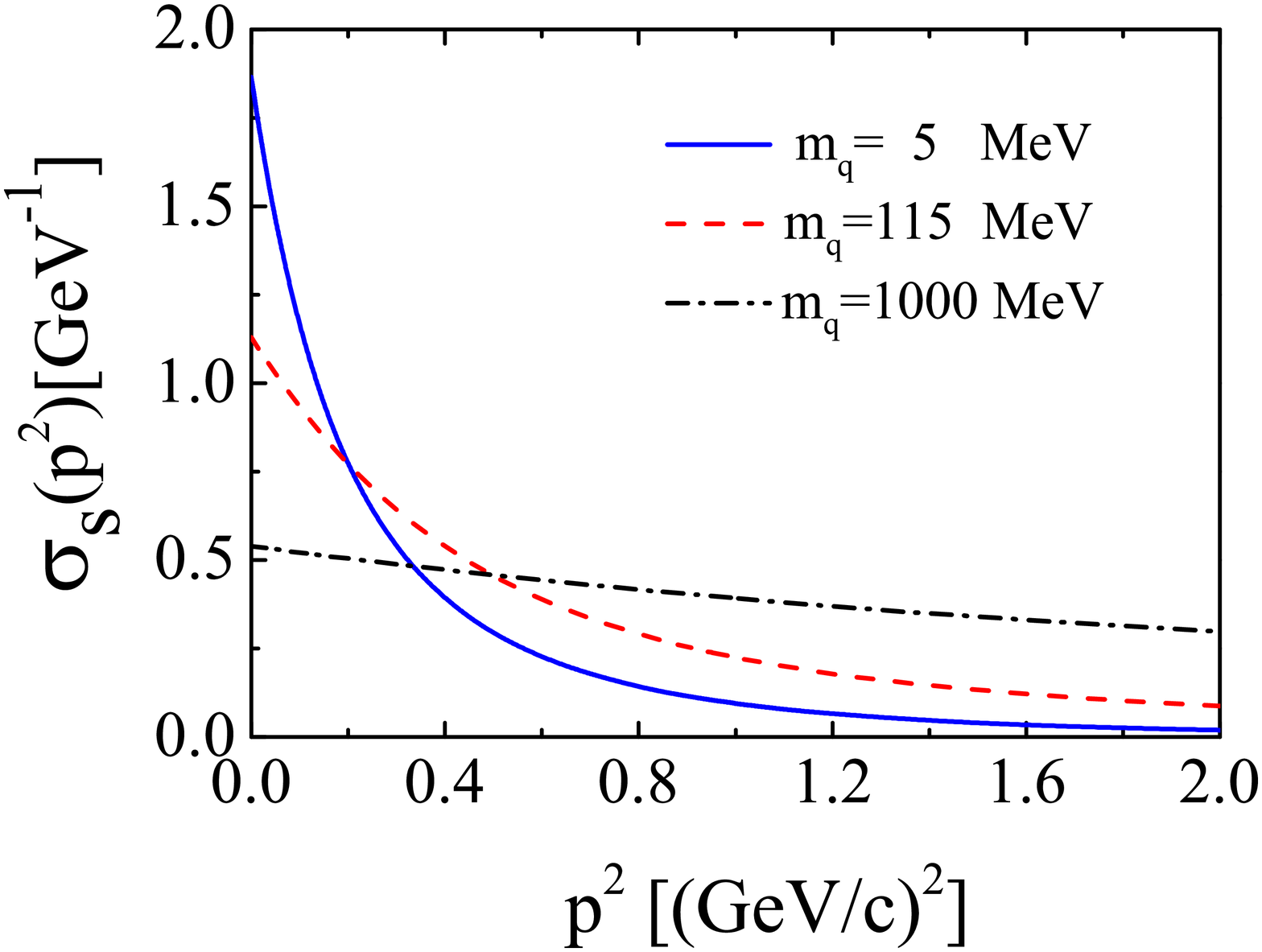}
\includegraphics[scale=0.25 ,angle=0]{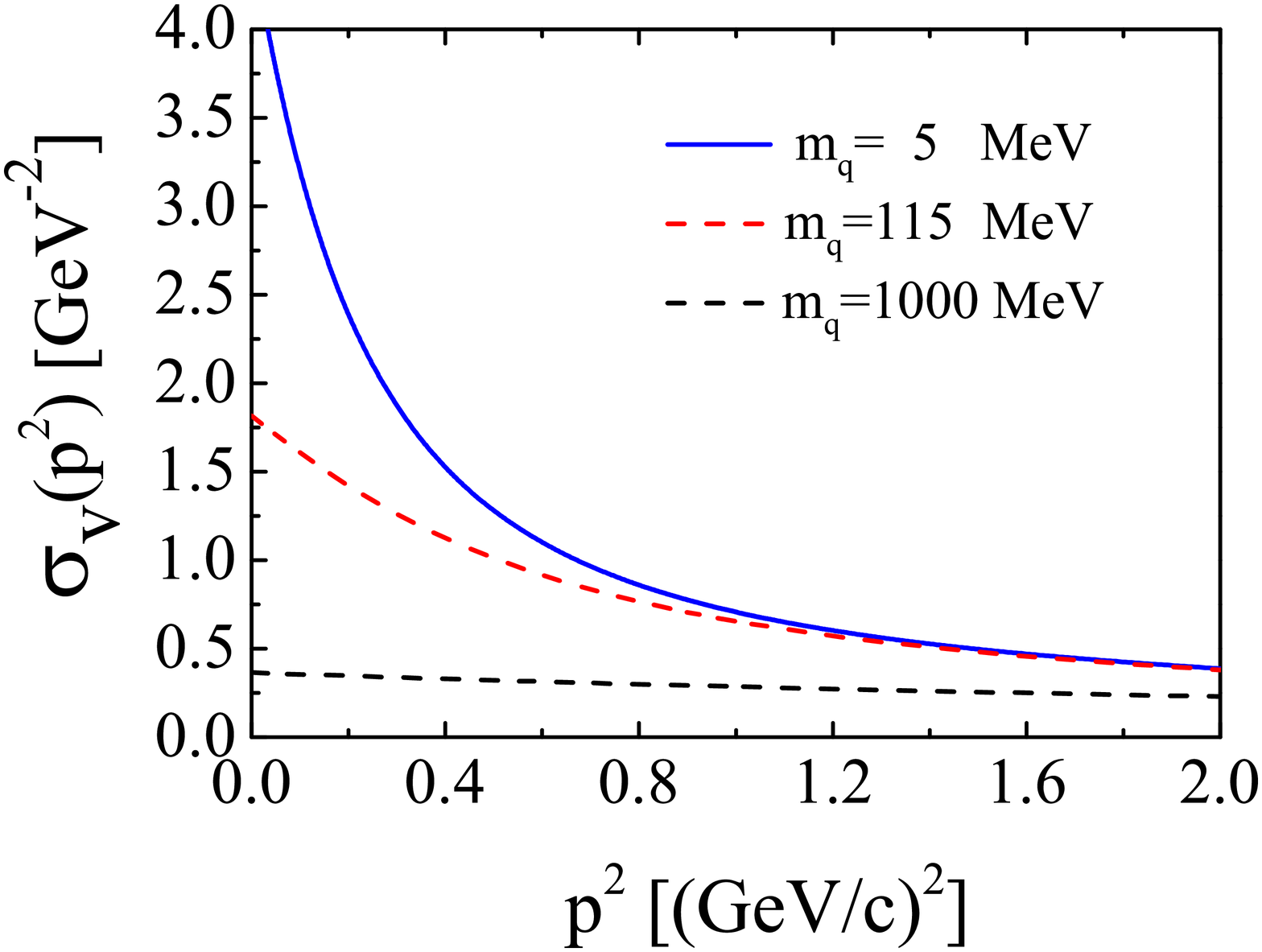}
\caption{(Color online) The solution of the tDS equations (\ref{a}) and (\ref{ab1})
with the IR part of the interaction kernel (\ref{phenvf}) only  as a function of the space-like
($p^2>0$)  real Euclidean momentum $p^2$.
 Solid curves correspond to $u$, $d$ quarks
with bare mass $m_q=5$~MeV, dashed curves depict results for   $s$ quarks with bare mass
$m_q=115$~MeV, and
dash-dotted curves  are for $c$ quarks with $m_q=1$ GeV.}
\label{fig1}
\end{figure}

In our subsequent analysis we
employ the effective parameters
from Refs.~\cite{Alkofer,Roberts}, $\omega=0.5$ GeV and $D=16 \, {\rm GeV}^{-2}$.
Results are  shown in Fig.~\ref{fig1} as   momentum dependence of the functions  $A(p),B(p)$ and $ \sigma_{s,v}(p)$
for different   bare masses:
${m}_q=0.005$ GeV for $u$,  $d$, ${m}_q = 0.115$ GeV  for $s$,
and ${m}_q=1$ GeV for $c$ quarks.
 It is seen that all these quantities are smooth, positively defined functions not containing any singularity
 along the space-like   Euclidean momentum $p^2$. The solution then is generalized to
 complex values of $p^2$ needed  to solve the BS equation for bound states.
 As mentioned above, such  solutions provide
 a good description of  many properties of light mesons (masses, widths, decay rates etc., cf.~\cite{Maris:2003vk,Holl:2004fr,Jarecke:2002xd,Krassnigg:2004if,Roberts:1994hh,Maris:2000sk,Maris:1999bh,ourFB}).
 However, an attempt to apply these solutions for heavier
 mesons with at least one light quark, e.g. for open charm $D$ mesons,
  leads to instabilities   in the numerical  procedure of solving the BS equation\footnote{In Ref.~\cite{SouchlasBS}
  the situation is described by the phrase "the ladder-rainbow model kernel
  has deficiencies in the heavy quark region that are masked in $QQ$ mesons but are plainly evident in
  $qQ$ mesons". If one attributes "deficiencies" with singularities in the propagator functions, it
  is the light-quark propagator which causes the problems in composites at energy scales $\gtrsim 1$ GeV.}.
  Obviously, this  is due to the fact that the integration
  domain   for heavier mesons becomes larger, cf. right panel of Fig.~\ref{parabola}, and the
  singularities in the propagator functions approach closely or  even  intrude into
  the corresponding parabola. Hence, a more detailed investigation of the
  behaviour of the propagator function in the complex Euclidean plane is required.

\section{Solutions of the \lowercase{t}DS equation in complex plane}
\label{sec3}
 As evident  from Fig.~\ref{parabola}, the parabolic integration domain for solving
 the BS equation can be conveniently divided into two parts: (i) one
(infinite) region where Re\,$p^2 > 0$, and (ii) a second one where Re\,$p^2 < 0$, which is
restricted  by the meson mass $M_{q\bar q}$, i.e. with the minimum (negative) value
Re\,$p^2 =- {M_{q\bar q}^2}/{4}$.
\subsection{Solutions along rays $\bphi=\bf const$}
\label{subsecphi}
From   eqs.~(\ref{a}) and (\ref{ab1}) it is explicitly seen that, in the right
hemisphere,  the integrals in the DS equation converge.
That means, an analysis of the behaviour of the solution for Re\,$p^2>0$ for large $|p^2|$ can be accomplished  directly
by means of  Eqs.~(\ref{a}) and (\ref{ab1}), i.e.   one can perform a rotation of the real axis by an angle
$\phi_p < \pi/4$ (for $p^2$, the rotation angle corresponds to $\phi_{p^2}< \pi/2$
thus covering the full half plane Re\,$p^2>0$) and
 solve the tDS equation for momenta $p=|p|\, \exp(i\phi_p)$ along    rays $\phi_p=$~const. In such a way one can obtain
 solutions of DS equation in the whole right hemisphere, including  large values of momenta, $|p^2|\to \infty$.
That method turns out to be  extremely efficient. The iteration procedure
converges rather fast and allows for a detailed analysis of  the
 solutions  $A(p)$, $B(p)$, $\sigma_s(p)$ and $\sigma_v(p)$ in a large interval   Re\,$p^2>0$ in the complex plane.

\begin{figure}[!ht]
\includegraphics[scale=0.25 ,angle=0]{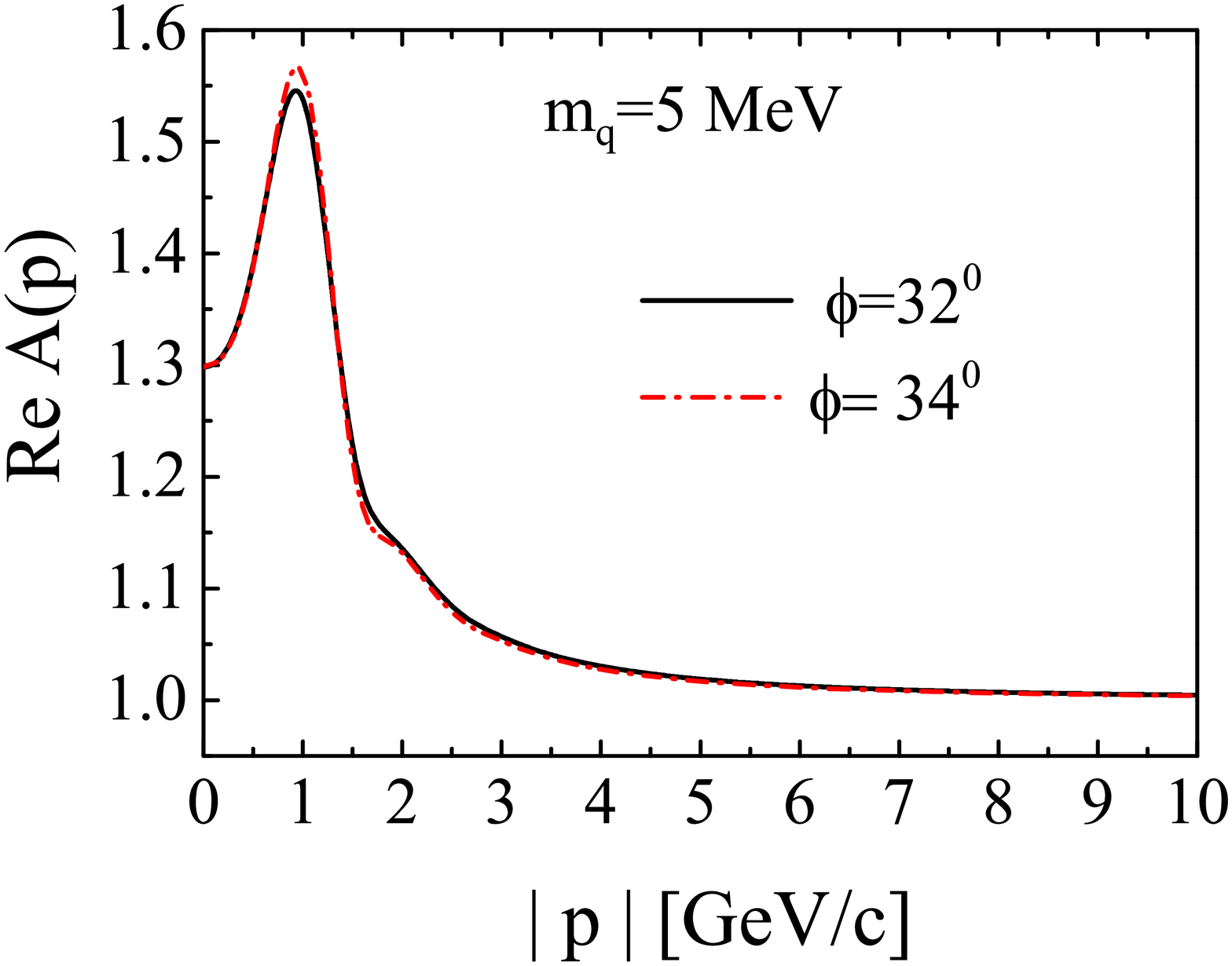}
\includegraphics[scale=0.25 ,angle=0]{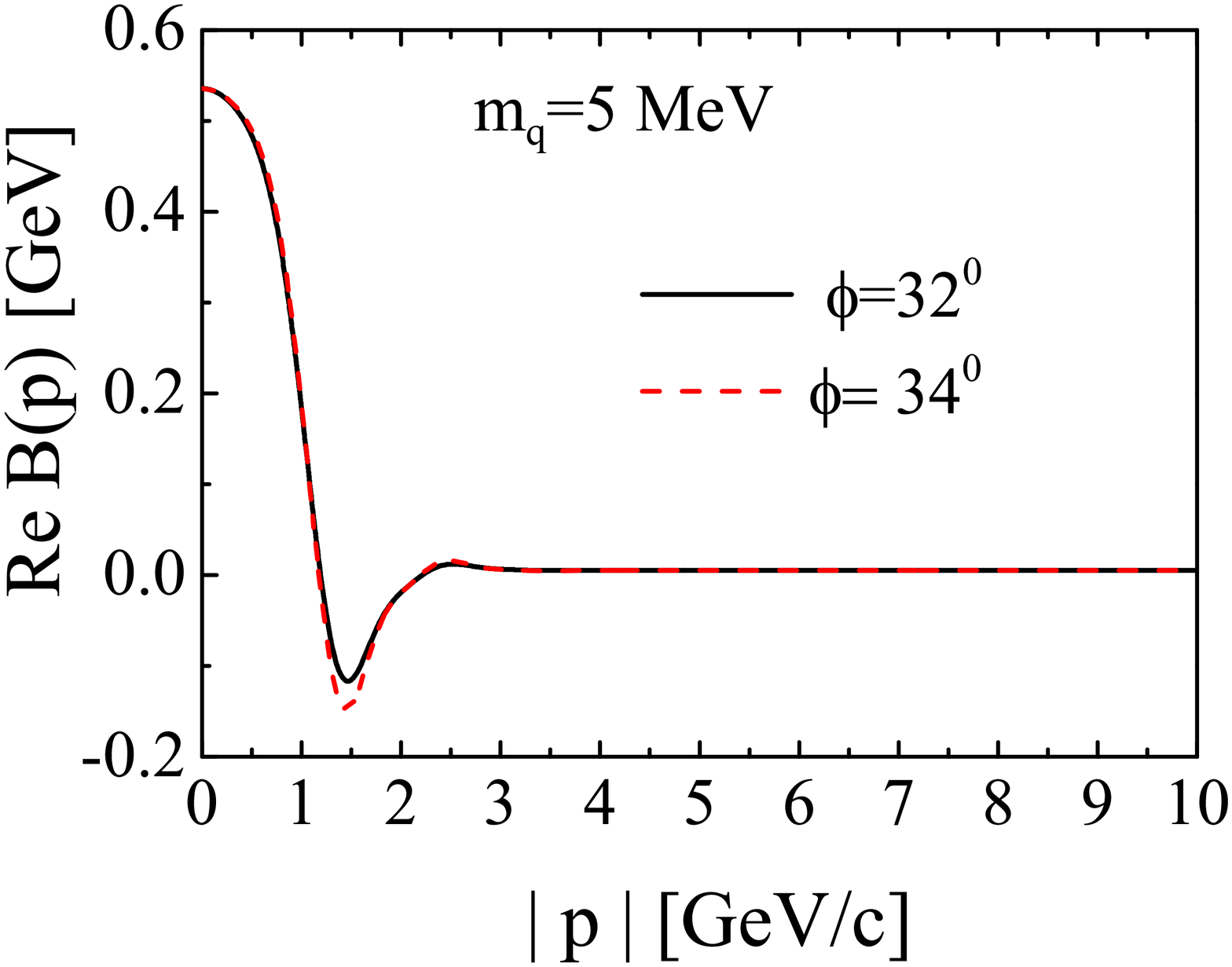}
\includegraphics[scale=0.25 ,angle=0]{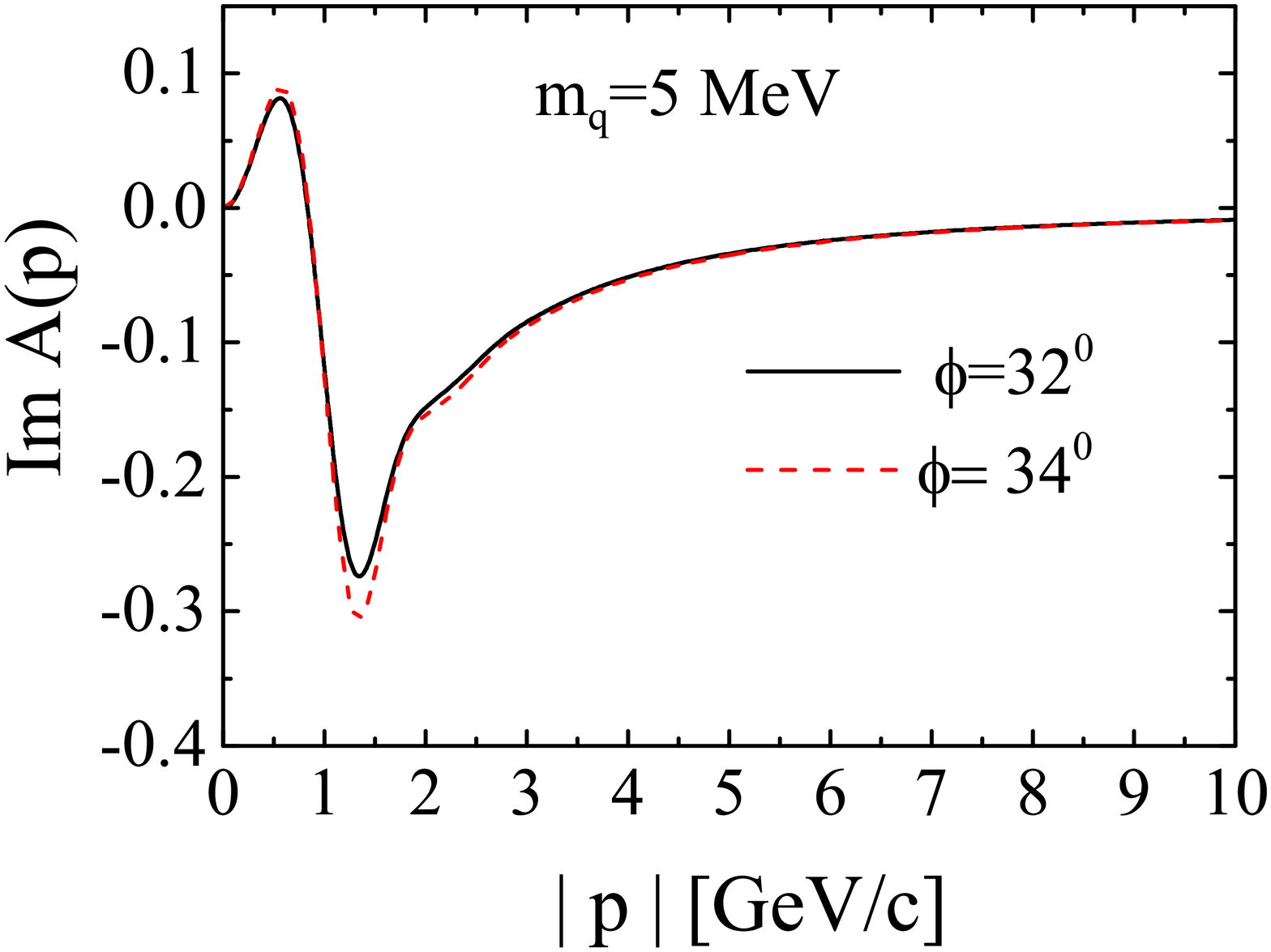}
\includegraphics[scale=0.25 ,angle=0]{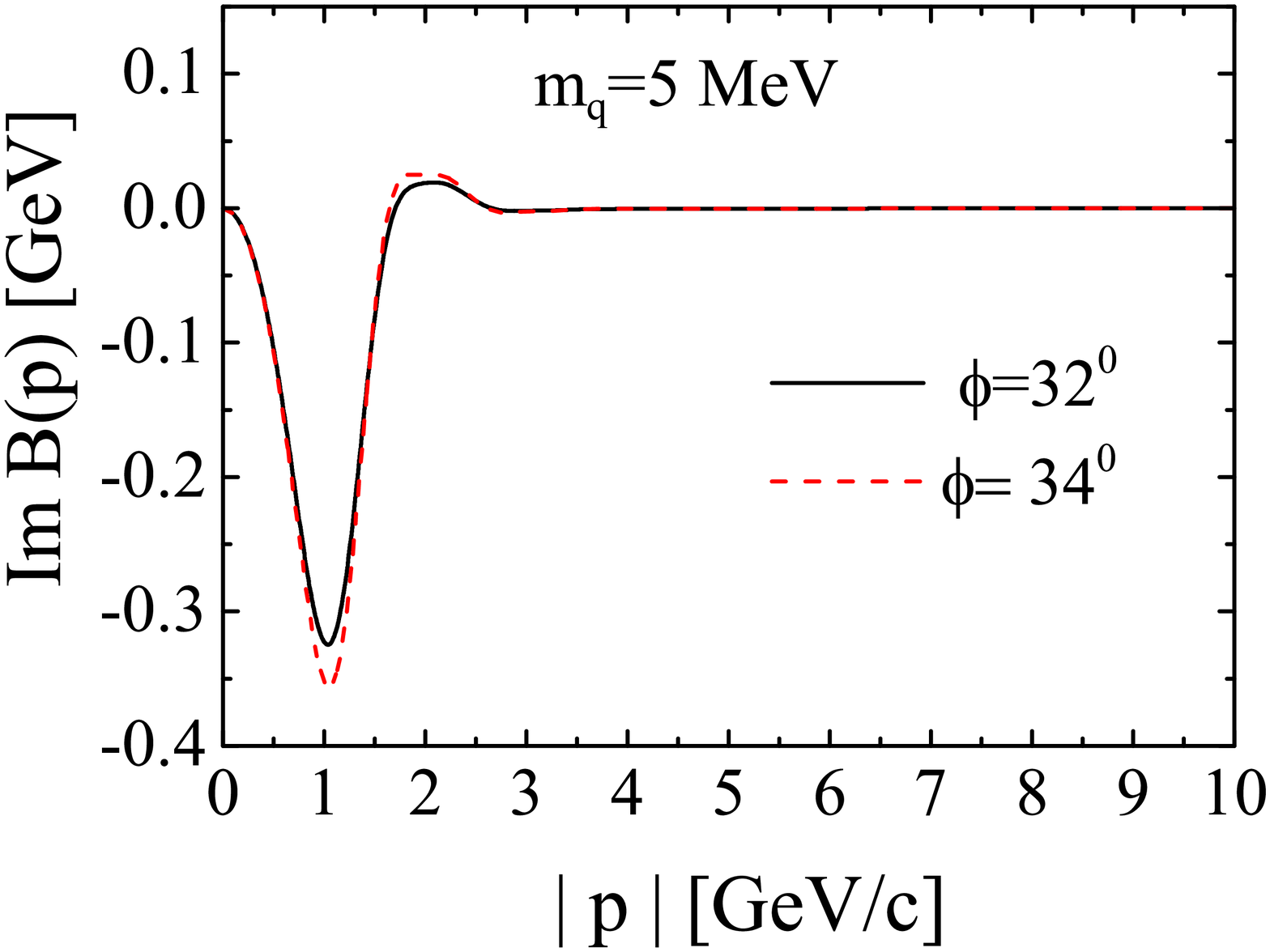}
\caption{(Color online) The solution of the tDS equations (\ref{a}) and (\ref{ab1}) in the
complex Euclidean space Re\,$p^2>0$ along rays $\phi_p=$~const  for two values of $\phi_p$
(black solid curves: $\phi_p=32^o$, red dashed curves: $\phi_p=34^o$)   for $u$, $d$ quarks with
$m_q=5\, MeV$.}
\label{reim}
\end{figure}

 In Fig.~\ref{reim}, we present, as an example, the solutions $A(p)$ and $B(p)$ in the complex
 plane along rays $\phi_p=const$ for two adjacent values of $\phi_p$. The
  complex solutions are smooth and smoothly changing under
  variations of $\phi_p$ as long as $\phi_p < \pi/2$, however, not anymore positively definite  so that,  in principle, some
 combinations of $A(p)$ and $B(p)$, in particular the inverse propagator part
 $\Pi(p^2)= p^2 A^2(p)+B^2(p)$, can vanish at certain values of $p^2$. This would imply the
 existence of pole-like singularities  for the propagator  functions $\sigma_{s,v}(p^2)$.
 To check this we   compute  integrals of  $\Pi(p^2)$  along wedge-shaped  closed contours  formed  by
 two rays with $\phi_p^{(1)}$ and $\phi_p^{(2)}$ and an
 enclosing curve provided by a section  of a circle with a radius  $|p_0|$. In what follows, such integrals
 along closed contours are referred   to as the  Cauchy integrals.
 A non-zero value of the Cauchy integrals will signal the occurrence of
  singularities  within the chosen contour. We calculate  the Cauchy integrals
 for many choices of integration contours and
 find them vanishing, i.e. the  functions $A$ and $B$ are analytical.
 Moreover, we find also that the propagator functions $\sigma_{s,v}(p^2)$ are analytical as well.
 Then, evidently  all singularities of the solutions $A$, $B$  and the
 propagator functions $\sigma_{s,v}$, if any, are to  be searched for
 in the left hemisphere, i.e. at  $ Re \ p^2 < 0$.
  It should be pointed out that an analogous rotation of axes  to
  this region, where $\phi_{p^2}>\pi/2$,
  is impossible since, as seen from Eqs.~(\ref{a}) and (\ref{ab1}), at   arguments larger than  $\pi/2$,
  negative values of Re\,$p^2 < 0$ lead to  divergent integrals  as $|p^2| \to \infty$.
 Hence, simple  analytical  continuation of the tDS equation along rays $\phi_p=$~const in the whole complex plane is impossible.

\subsection{Solution of the \lowercase{t}DS equation at ${\mbox{\boldmath$ Re\ p^2 <0$}} $}
\label{subsectimelike}

 In our calculations  we are interested in a restricted   domain  in  the
 left hemisphere  for which $|p^2|$ is relatively small, Re\,$p^2> -M_{q\bar q}^2/4$
 where the tDS integrals converge.
 That means, at such values of   $|p^2|$   the tDS equation  along the positive real axis
 can still be used  to find the solution in the complex plane.
 It is worth mentioning that, in principle, one does not need to solve Eqs.~(\ref{a}) and (\ref{ab1}) for each point inside
 the parabola (\ref{parequation}). It suffices to know whether the solutions $A(p^2)$ and $B(p^2)$  are analytical in the
 corresponding domain, and if so, one can find  the solutions $A(p^2)$ and $B(p^2)$  only once along the
 contour defined by the parabola and,
 due to the Cauchy's theorem,
\begin{eqnarray}
 A(z) = \frac{1}{2\pi i}\oint  \frac {A(\xi)}{\xi-z} \  d\xi,
\label{cauchy}
\end{eqnarray}
  to determine $A(p^2)$ and $B(p^2)$   in any other desired point, see  Ref.~\cite{fisher}.
Consequently, we proceed with an analysis of the domain of analyticity of $A(p^2)$, $B(p^2)$ and $\sigma_{s,v}(p^2)$
in the region $-M_{q\bar q}^2/4<Re\, p^2< 0$.

We begin with an analysis of properties of the propagator of the $c$ quark for
which, as we found in our previous calculations \cite{ourFB},
the procedure of solving the BS for $M_{q\bar q} <  2$ GeV is rather stable. This is a hint that  the solutions
$A(p^2)$ and  $B(p^2)$ and the propagator functions  $\sigma_{s,v}(p^2)$ could be  analytical functions everywhere
in the integration domain.
Indeed, we solve the tDS  equation for $A(p^2)$ and  $B(p^2)$ in a large interval of  Re\,$p^2 <0$,
compute the corresponding Cauchy integrals and find them to be always zero.
Moreover, we find that  $\sigma_s(p^2)$ and $\sigma_v(p^2)$ are also analytical
functions within the corresponding parabola.
This means that for such analytical functions one can find convenient parametrizations
in terms of rational functions, cf.~\cite{walsh}, to be used in solving the tBS equation.
Remind that, as seen from the tDS equation, the solutions $A(p^2)$ and $B(p^2)$ are self-conjugated
functions, i.e. the propagator functions $\sigma_{s,v}(p^2)$ must be real along the real axis of $p^2$. This
restricts the possible form of parametrizations by the condition
$F(p^2,\{\alpha_i\})=F(p^2,\{\alpha_i^*\})$, where $\{\alpha_i\}$ is the set of free parameters, to be found by
fitting the solution along the real axis, see e.g. Ref.~\cite{souchlas}.
\begin{figure}[!ht]
\includegraphics[scale=0.35 ,angle=0]{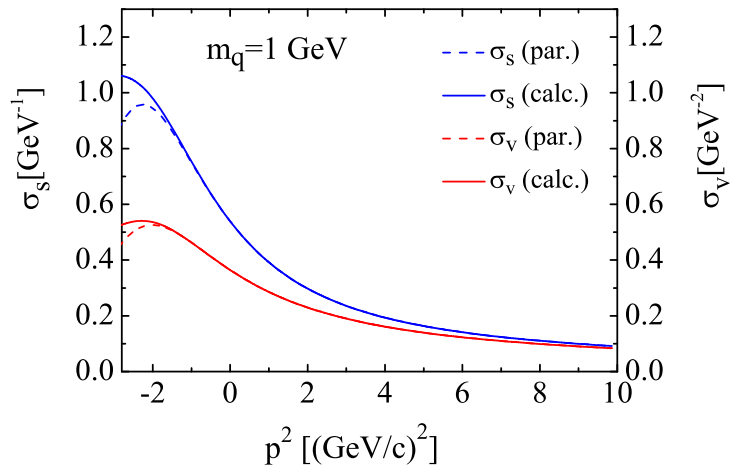}
\caption{(Color online) The functios  $\sigma_s$ and $\sigma_v$
 for real values of $p^2$. The  solid curves are for the numerically determined solutions,
while dashed curves depict the approximation (\ref{parx}) with $i=1,2$. For $m_q=1$ GeV.}
\label{sigmac}
\end{figure}
In Fig.~\ref{sigmac} we exhibit the behaviour of $\sigma_{s,v}(p^2)$
  along the real axis which are used in further parametrizations.  The propagator functions
are  smooth and obey a simple form which inspires   the following choice  for the parametrization
\begin{eqnarray}&&
\sigma_{s,v}(p^2)=\sum_i \frac{\alpha_i(s,v)}{p^2+\beta_i^2(s,v)} + \sum_i \frac{\alpha_i^*(s,v)}{p^2+\beta_i^{*2}(s,v)},
\label{parx}
\end{eqnarray}
where the complex parameters $\alpha_i$ and $\beta_i$ can be easily  obtained by fitting the corresponding
 solution along the real axis of $p^2$. We use the  Levenberg–-Marquardt algorithm for fitting.
We find that   for each function in Eq.~(\ref{parx}) two terms, i.e. eight parameters for each function,
are quite sufficient to obtain a good approximation of the solution.
In Table~\ref{tb1} we present   the sets of parameters $\alpha_{1,2}(s,v)$ and $\beta_{1,2}(s,v)$
obtained for $\sigma_s$ and $\sigma_v$, respectively,
from a fit in the interval $-1.5\, {\rm( GeV/c)} ^2< p^2 <10\, {\rm (GeV/c)}^2$.
The quality of the fit is demonstrated in Fig.~\ref{sigmac}, where the dashed lines represent the obtained
 approximation to the corresponding exact propagator functions (solid curves).
 In the interval $-1.5\, {\rm (GeV/c)} ^2< p^2 <10\, {\rm (GeV/c)}^2$, the achieved fit is excellent. Then we randomly
calculate the propagator functions at complex values of $p^2$  by solving numerically~(\ref{a}) and (\ref{ab1})
and compare with parametrizations~(\ref{parx}). The agreement is impressively good.
(It should be   noted, however that, since the  parametrized
functions are of a rather simple shape, the set of obtained parameters $\{ \alpha_i\}$ and $\{ \beta_i\}$
is far from being  unique, i.e.  one can achieve a similar quality of the fit with many other choices
of $\{ \alpha_i\}$ and $\{ \beta_i\}$. The only restriction is
that the "mass" parameters   $\{ \beta_i\}$ must not provide singularities, neither along the real axis, nor in the
complex plane inside the parabola, see also Ref.~\cite{Bhagwat:2002tx}.) In essence, Eq.~(\ref{parx})
with parameters in Table~\ref{tb1} provide a convenient parametrization of $\sigma_{s,v}(p^2)$ which is easily applicable
in the tBS equation for $M_{q\bar q} < 3$ GeV.
\begin{table}[!ht]
\caption{The parameters
 $\alpha_i(s,v)$ and  $\beta_i(s,v)$ for the effective parametrizations, Eq.~(\ref{parx}). For $c$ quark of mass $m_q=1$ GeV.}
\begin{tabular}{l c l lcc c c c } \hline\hline  
     j   &\phantom{p}   &  $\alpha_1(j)$ &\phantom{p} & $\beta_1(j)$ [GeV]           &\phantom{p}& $\alpha_2(j)$        &\phantom{p}&$ \beta_2(j)$ [GeV]        \\
         &\phantom{p}   &  (Re \, , \ Im )  &\phantom{p} &  (Re \, ,\ Im )   &\phantom{p}& (Re \, ,\ Im )  &\phantom{p}& (Re \, ,\ Im ) \\
         &              &                     &            &                     &           &                    &          & \\\hline
 $s$     &   & (1.409\, ,  0.9802)  [GeV] &    &(1.8627\, ,   0.602)              &  &(-0.909 \, ,   -0.245)  [GeV] &      & (1.875\,    0.886)     \\
         &   &                      &          &                     &           &                                &          & \\
 $v$     &   &(0.08624\, ,  0.598)   &    & (1.773\, , 0.7179)&           & (0.4145\, ,  -0.267)      &        & (2.112\, , 0.5177)    \\
                   &              &             &            &              &           &            \\
                     \hline\hline
\end{tabular}
\label{tb1}
\end{table}

 A different situation occurs when one tries to find  such  parametrizations for
 the propagator functions of light quarks. We find that, in spite of zero Cauchy integrals for
 $A(p^2)$ and $B(p^2)$, the propagator functions $\sigma_{s,v}(p^2)$ provide non-zero
  Cauchy integrals for $M_{q\bar q}> 1$ GeV. Besides, the values of the Cauchy integrals for
 $\sigma_{s,v}(p^2)$, i.e. their residues, are found to  depend  on the chosen
  contour inside or near the parabola. This  is a clear indication that
  $\sigma_{s,v}(p^2)$ have poles in this region and, moreover, the number of poles is greater than one.
 As noted above, for numerical calculations it is extremely important  to find, with a good accuracy,
   the position of the poles and the corresponding residues for   $\sigma_{s}(p^2)$ and   $\sigma_{v}(p^2)$.
In such a case, if the  complex valued functions $\sigma_{s,v}(p^2)$
  have only isolated poles $p_{0i}^2$ within a certain domain and are analytical on its closing contour $\gamma$, they can be represented as
 \begin{equation}
 \sigma_{s,v}(p^2) = \widetilde\sigma_{s,v}(p^2) + \sum_i\frac{{\rm res} [\sigma_{s,v}(p^2_{0i})]}{p^2-p^2_{0i}},
\label{main}
 \end{equation}
 where $\widetilde\sigma_{s,v}(p^2)$ are analytical functions within
 the considered domain and, consequently, can be computed as
 \begin{equation}
 \widetilde\sigma_{s,v}(p^2)
  =\frac{1}{2\pi i}\oint\limits_{\gamma}\frac{\widetilde \sigma_{s,v}(\xi)}{\xi -p^2} d\xi
 =  \frac{1}{2\pi i}\oint\limits_{\gamma}\frac{\sigma_{s,v}(\xi)}{\xi -p^2} d\xi .
  \label{f1}
 \end{equation}
Then, Eqs.~(\ref{main}) and (\ref{f1}) imply that, in solving the BS equation, the first term in
(\ref{main}) is free of singularities and does not require modifications of the numerical procedure;
the second term also does not cause numerical troubles
since integrations  can be performed analytically. A similar strategy in solving numerically the tBS equation
in presence of poles has been proposed in Ref.~\cite{Blank:2011ha}.

\section{The pole structure of solutions of the \lowercase{t}DS equation}
\label{polestruc}
In order to be able to use the representation (\ref{main}) on needs to know  the
positions of poles and the corresponding residues of the propagator functions.
To this end  we solve the decomposed tDS equations (\ref{a}) and (\ref{ab1}) for $A(p^2)$ and $B(p^2)$ and investigate the
above mentioned scalar part $\Pi (p^2)= p^2A^2(p^2)  + B^2(p^2)$  of the  quark propagators
$S(p^2)$ together with the  functions
$\sigma_{s,v}(p^2)$ by computing Cauchy integrals along
closed contours. If the corresponding Cauchy integral for $\Pi (p^2)$  is zero and for $\sigma_{s,v}(p^2)$ finite,
this immediately implies that $\Pi (p^2)$ has zeros and  $\sigma_{s,v}(p^2)$ have poles inside the considered region.
\subsection{Searching for singularities}

To find the appropriate contour enclosing zeros of $\Pi (p^2)$   we proceed as follows.
The complex function $\Pi (p^2)$ is presented as a vector field in the complex momentum
plane Re\,$p{\rm -Im}\,p$ in a certain region of $p^2$.
A vector field has a smooth distribution of its force lines if it does not contain null vectors
(see e.g Ref. \cite{krasnoselsky}).  Contrarily,
in the vicinity of null vectors (for the vector field these  are known as singular vectors) the
force lines exhibit a vortex-type behavior. This can essentially facilitate our analysis since a contour around
the vortices  of $\Pi (p^2)$ definitely   contains zeros.

\begin{figure}[!ht]\hspace*{-12mm}
\includegraphics[scale=0.38 ,angle=0]{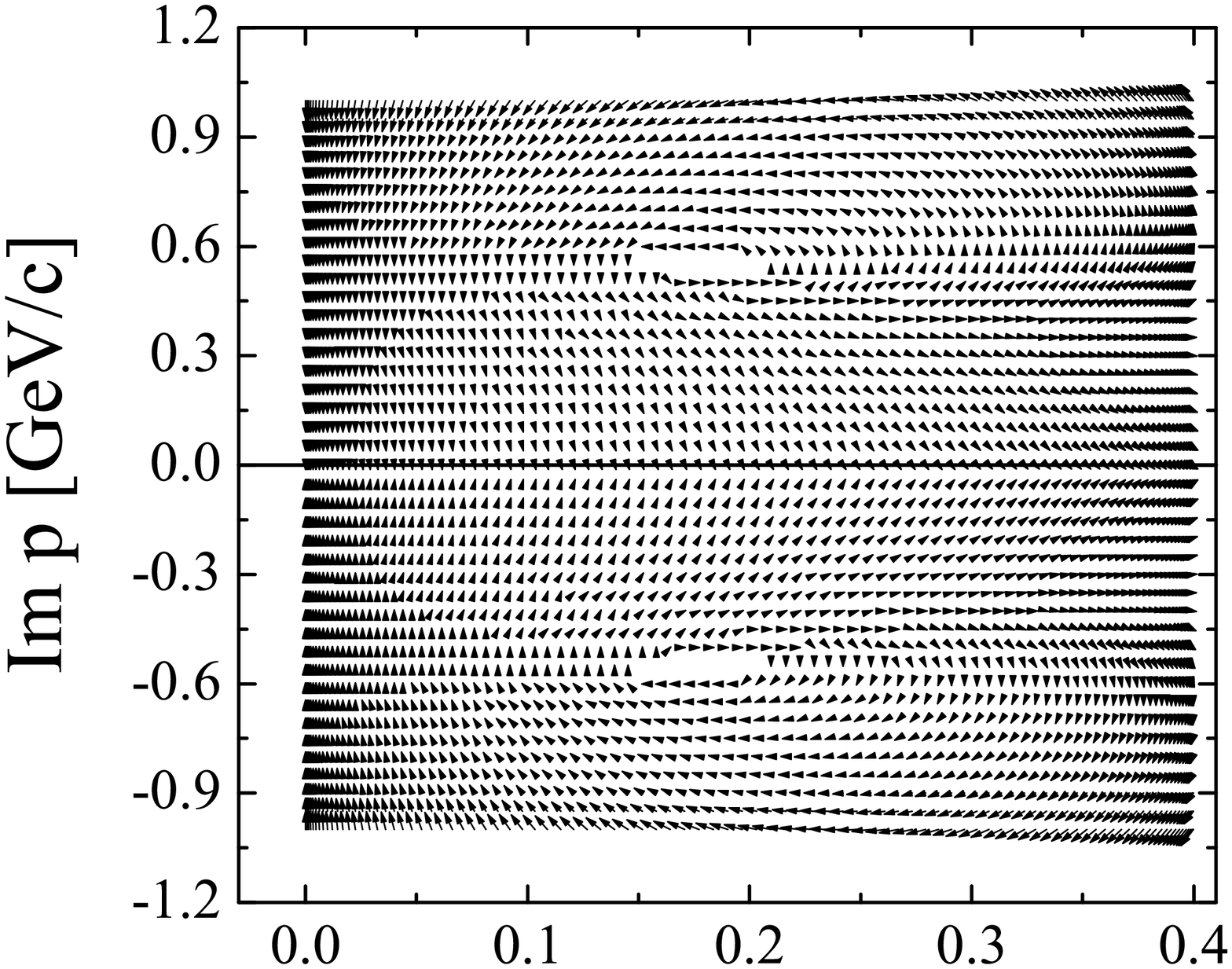} \hspace*{6mm}
\includegraphics[scale=0.36 ,angle=0]{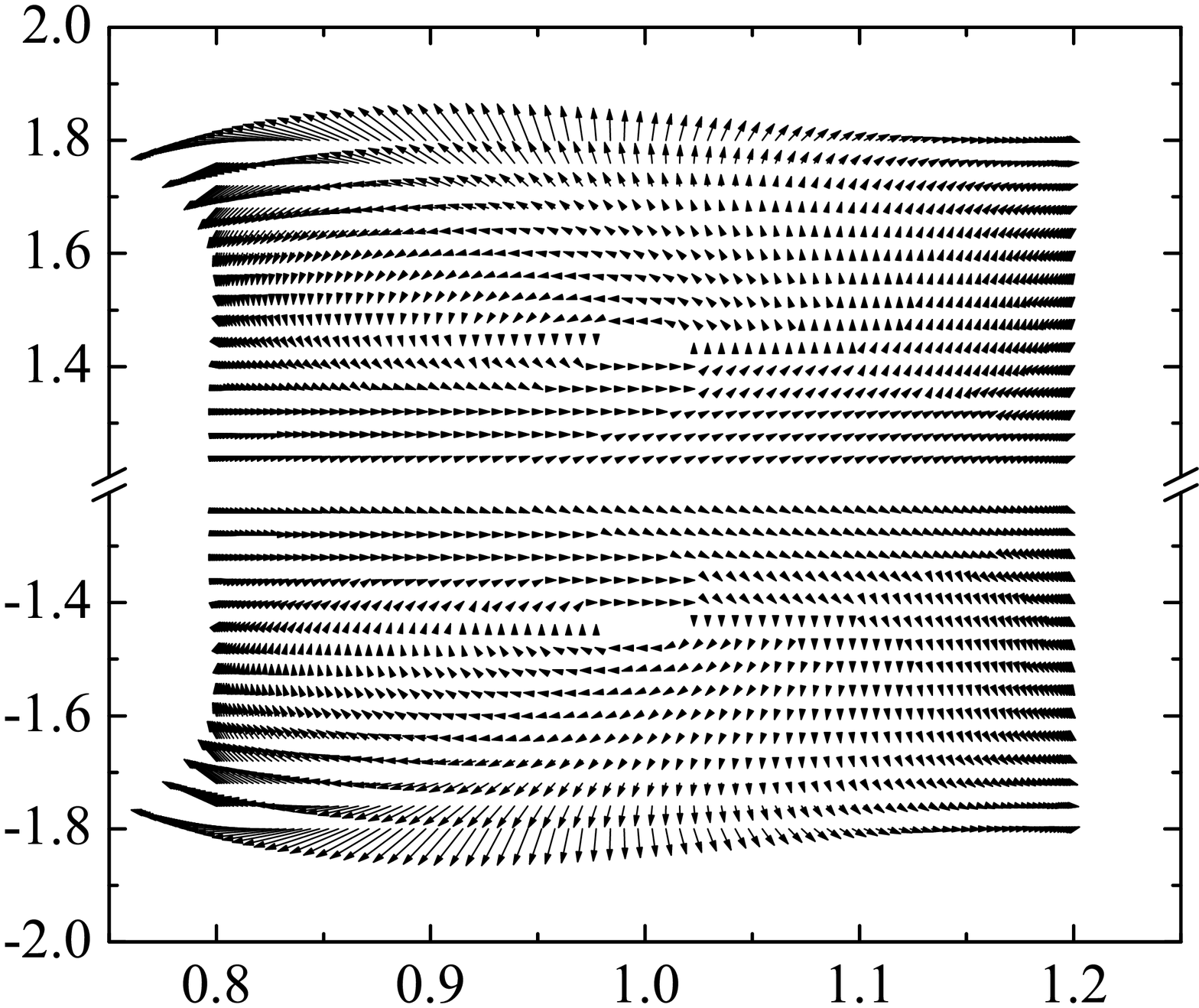} \hspace*{-10mm}\vspace*{5mm}
\includegraphics[scale=0.38,angle=0]{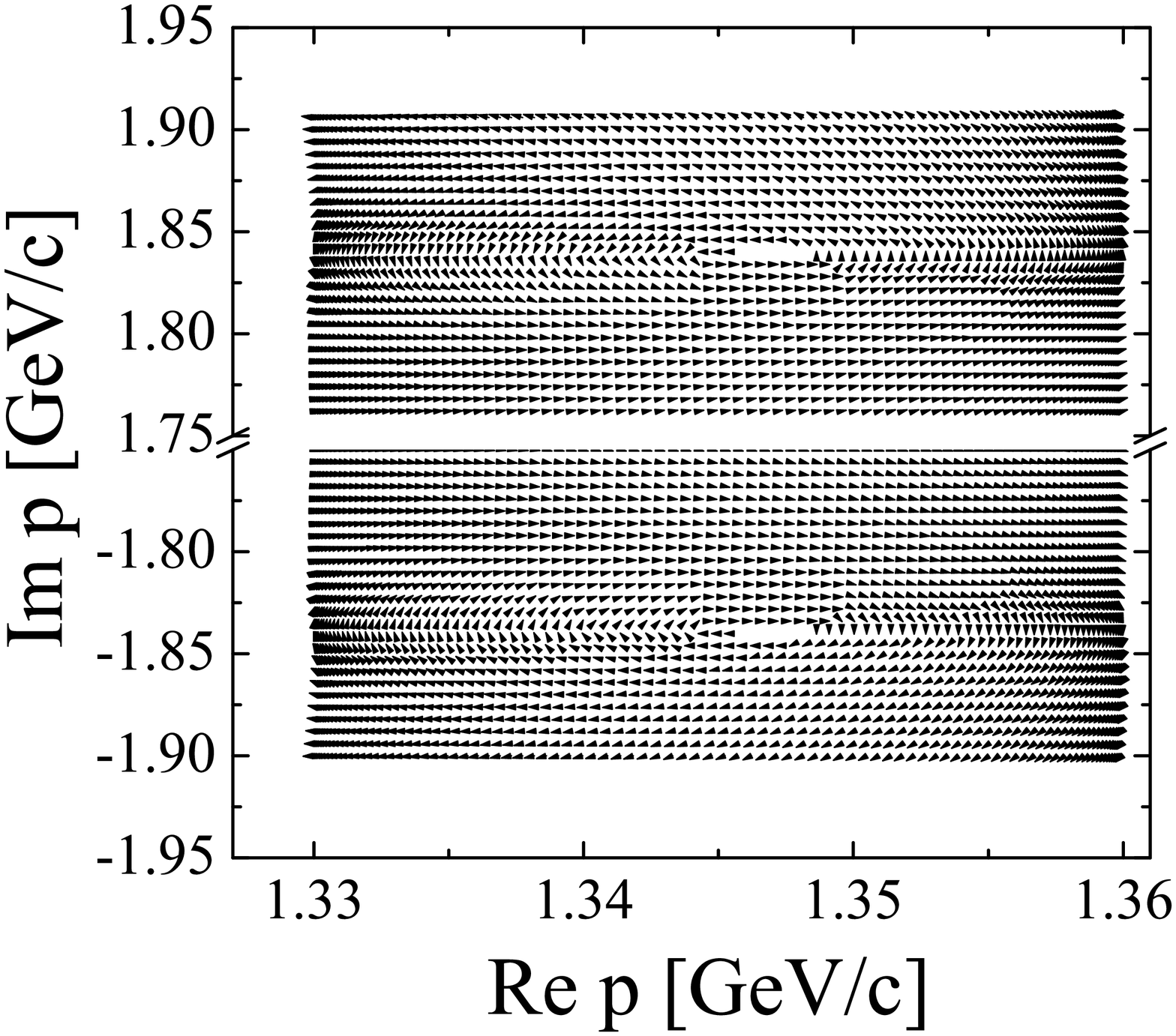} \hspace*{3mm}
\includegraphics[scale=0.38 ,angle=0]{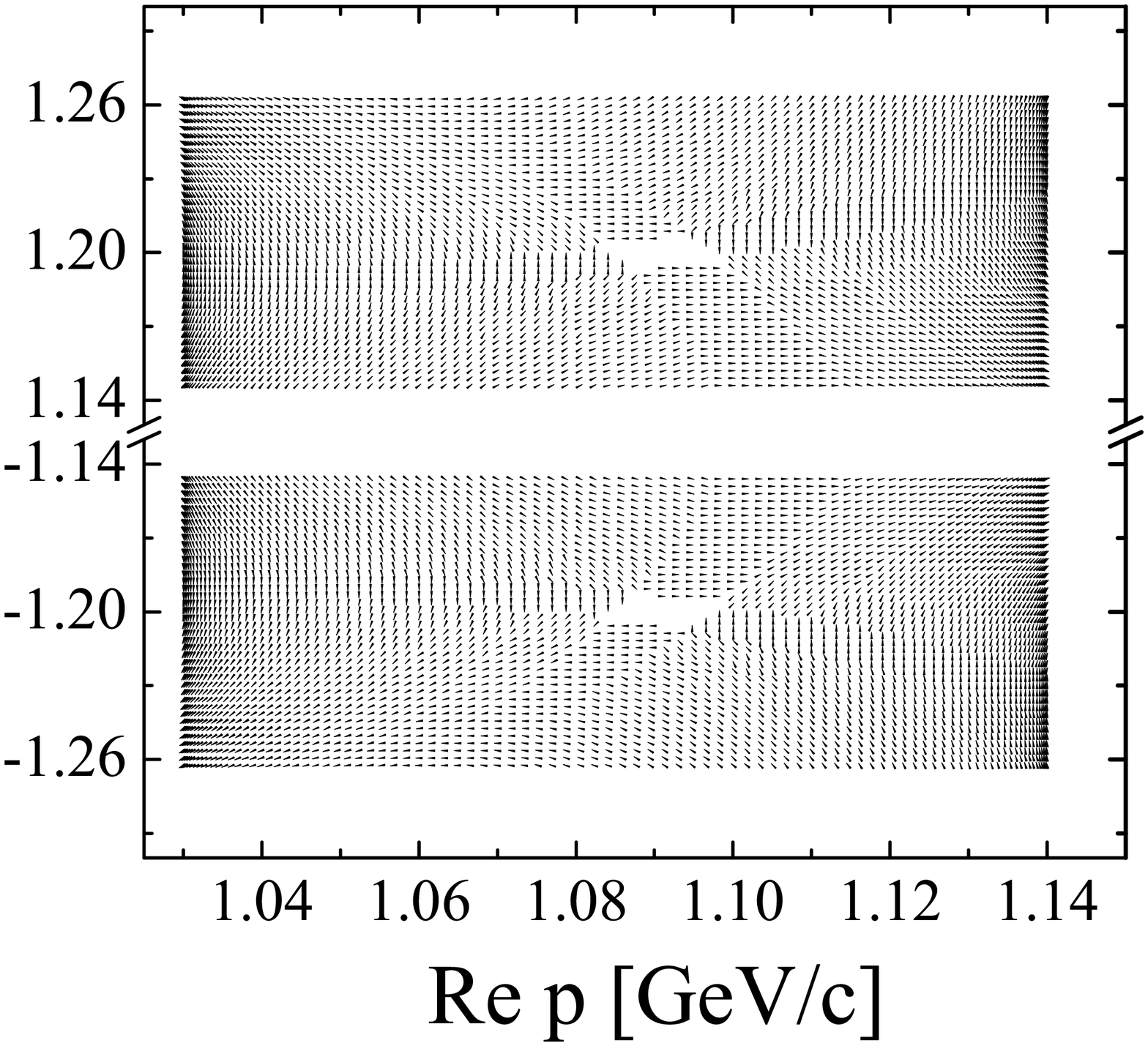}
\caption{ (Color online) The   force lines  of the inverse propagator part $\Pi(p^2)$ for  $u$, $d$ quarks.
 The self-conjugated nature of the
singularities located within the  "turbulent" regions is evident.}
\label{poliusa}
\end{figure}

In Fig.~\ref{poliusa} we present a few selected regions of the
Re\,$p{\rm -Im}\,p$ plane, where the vector field of  $\Pi (p^2)$ has been found to exhibit
 vortex-type structures.
Clearly, the zeros and singularities must be searched for in the vicinity of these vortices.
The force line method, as other methods such as contour plots or three dimensional visualizations,
 are useful for surveys.
\subsection{Allocation of singularities}
The further strategy of finding the positions  of singularities more accurately  is as follows:\\
(i) Choose a vortex, enclose it  with a contour and compute the Cauchy integrals
 of $A(p^2)$, $B(p^2)$ and $\Pi(p^2)$.
  Vanishing integrals imply that these functions are analytical within the chosen contour.\\
(ii) Compute the Rouch\'e's  integral\footnote{Rouch\' e's integral of an analytical
complex function $f(z)$ on a closed contour $\gamma$  is defined as
$\frac{1}{2\pi i}\oint\limits_\gamma\frac{f'(z)}{f(z)}d z$.} of the function $\Pi(p^2)$. Since in the previous item   we found
 $\Pi(p^2)$ to be  analytical, such an integral, according to the Rouch\'e's theorem,  gives exactly the number
  of its zeros  inside the  contour.\\
(iii) Compute the Cauchy integral of the propagator functions $\sigma_{s,v}(p^2)$  which,
   if the  Rouch\'e's integral is found to be   an integer positive
number, clearly must be different from zero.
 Moreover, the non-zero Cauchy integrals provide us with the residues needed in Eq.~(\ref{main}).
To find the desired  singularities  we compute the following integrals

 \begin{eqnarray}&&
 \oint\limits_\gamma \left[ \xi^2 A^2(\xi))+B^2(\xi)\right ] d \xi^2 =0,  \label{ucl}\\ &&
 \frac{1}{2\pi i} \oint\limits_\gamma
\frac{  \left[ \xi^2 A^2(\xi))+B^2(\xi) \right]'_{\xi^2} }{\xi^2 A^2(\xi)+B^2(\xi) }d\xi^2 =N_z, \\[1mm] &&
 \frac{1}{2\pi i}\oint\limits_\gamma \sigma_{s(v)}(\xi^2) d\xi^2 =
 \sum_i res[\sigma_{s(v)}(\xi_i^2)] \label{resids}
 \label{residv}
 \end{eqnarray}
(an effective algorithm
for numerical evaluations of Cauchy-like integrals can be found in, e.g. Ref.~\cite{ioak}).
 Then we squeeze the contour around the chosen vortex while  holding the   conditions (\ref{ucl})-(\ref{residv})
 until the desired accuracy of the determination of the pole position is attained.
 In such a way we find the first few poles of $\sigma_{s,v}(p^2)$ together with their residues relevant
 for~(\ref{main}).
 Note that a good numerical test of the performed calculations is the following procedure.
  Enclose a few vortices by a larger contour and
  ensure that  Rouch\'e's integral is exactly equal to the number of vortices and the Cauchy integrals of
 $\sigma_{s,v}(p^2)$ are exactly the sum of individual residues found before for each isolated vortex.
 The technique of computing Cauchy integrals to determine the analyticity of the propagator functions has been exploited
also in Ref.~\cite{holties} within a simplified  kernel with
the  running coupling replaced by a constant and the IR term taken as $\sim 1/q^4$.
\begin{table}[!ht]
\caption{The pole structure of  the propagator
functions for   $u/d$ and $s$ quarks.  The pole positions $(Re\ p^2, Im\ p^2)$ are in
units of  ${\rm (GeV/c)}^2$, while the residues of $\sigma_s$ are in GeV;
residues of  $\sigma_v$ are dimensionless.   Only the first, four self-conjugated poles on $p^2$
close to the parabolas in   Fig.~\protect\ref{parabola} are presented.}
\begin{tabular}{lccccc} \hline
        $u$, $d$ quarks            &             1                  &        2                  &            3                                &  4  \\ \hline
  pole position     &  (-0.2588, $\pm$ 0.19618)        &  (-0.2418, $\pm$ 2.597)   &      (-1.0415,$\pm$    2.8535)   &(-0.738,0.0)\\
 res[$\sigma_s$ ]   & (-0.016, $\mp$ 0.511)         & (0.04,$\pm$ 0.10)      &      (-0.05, $\mp$    0.076)    & (0.069,0.0)\\
 res[$\sigma_v$]    & (0.259,   $\mp$  0.859)         & (0.0234,$\mp$ 0.063)   &  (0.0014, $\mp$ 0.052 )       & (-0.080,0.0)
 \\ \hline
  $s$ quarks        &             1               &        2              &            3                                &  4  \\ \hline
  pole position     & (-0.436, $\pm$ 0.513)       &  (-0.51,$\pm$ 3.35)   &      (-1.45, $\pm$  3.82)       &(-3.25,0.0)\\
 res[$\sigma_s$ ]   & (0.009, $\mp$ 0.49)         & (0.06,  $\pm$ 0.10)   &      (-0.056,$\mp$ 0.08)        & (0.007,0.0)\\
 res[$\sigma_v$]    & (0.26,  $\mp$ 0.54)         & (0.013, $\mp$ 0.06)   &     (-0.0005,$\mp$ -0.048 )     & (0.004 ,0.0)
 \\ \hline\hline
\end{tabular}
\label{tb2}
\end{table}

\begin{figure}[!ht]
\includegraphics[scale=0.7 ,angle=0]{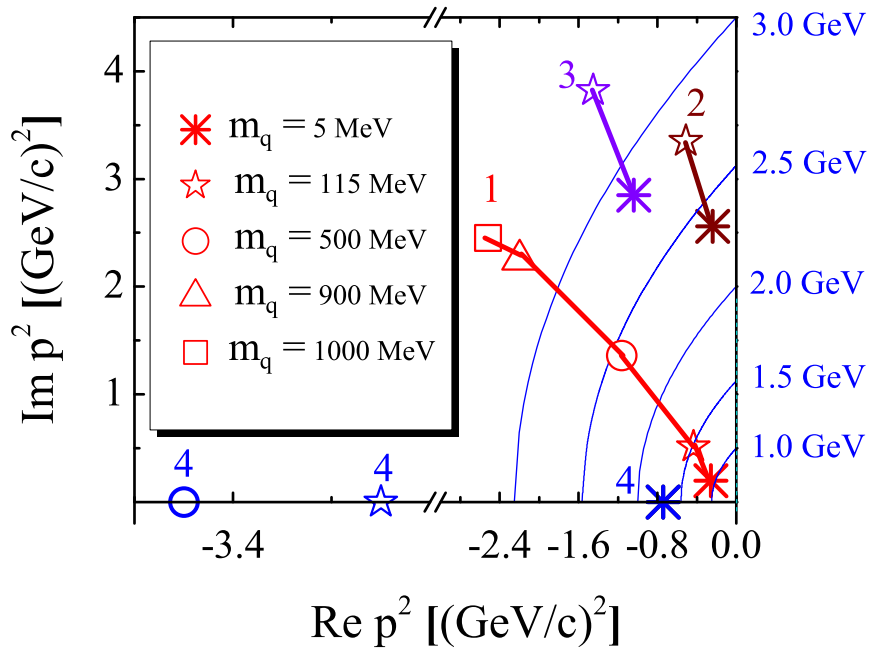}
\caption{Positions of few first poles in the upper hemisphere of the complex $p^2$ plane,
 labeled in correspondence to  the Table~\ref{tb2},  for $u$, $d$ (asterisks) and
 $s$ (open stars) quarks. The relevant sections of the parabola (\ref{parequation})  corresponding to
 the meson bound-state mass $M_{q\bar q}$
 are presented for $M_{q\bar q}=1, \, 1.5,\, 2,\, 2.5$ and $3$  GeV, from right to left.
 To emphasize the dependence of the pole positions  on
 the bare quark mass $m_q$, the "lowest" poles for the fiducial values $m_q=0.5$  GeV (open circle),
  $m_q=0.9$ GeV (triangle) and $m_q=1$ GeV (open square) are displayed as well. As an illustration
  of the behaviour on the real axis,
  the fourth pole for $m_q=0.5$ GeV is presented (circle).
  The tendency is    that, with increasing  bare quark mass, the corresponding
pole is shifted towards larger values of $Im\ p^2$ and   $|Re\ p^2|$. The area to the left after the axis
break is the "unrevealed terrain", where further singularities could be located.}
\label{rectangular}
\end{figure}

  Results of our calculations are presented in Table~\ref{tb2} and Figs.~\ref{parabola}~and~\ref{rectangular}.
From  Table~\ref{tb2} and Figs.~\ref{parabola} and \ref{rectangular} it is seen that for
$M_{q\bar q}< 1$ GeV all singularities in  the light
quark propagator are located outside the parabola.
This allows to establish easily reliable
algorithms for  solving  numerically the tBS equation~\cite{ourFB} for such a case.
For larger meson masses, e.g.  $M_{q\bar q}\sim 2$ GeV, the
singularities are either within the tBS domain (\ref{parequation}), or in the close vicinity
  (see Fig.~\ref{parabola}, right panel) and,
as a result, the numerical procedure adequate for low masses requires a proper modification.
We can take advantage of the fact that,  for meson masses
$0.14 {rm GeV}< M_{q\bar q}\ \lesssim 2~{\rm GeV}$, the number of  singularities is not too
 large and one can combine Eqs.~(\ref{parx}) and (\eqref{main})  in such a manner that
 the regular part $\widetilde \sigma_{s,v}(p^2)$ can be used as above
 without modifying the algorithm. The pole part, however,
 for heavier quarks allows to carry out
the angular integration analytically by employing  parametrizations (\ref{parx}).
Note that the propagator functions for the $s$ quark ($m_q=115\, MeV$) possess only a self conjugated
pole in the vicinity of the considered parabola, cf.~Fig.~\ref{parabola}.
It is located at $p^2=(-0.436 \pm  0.5131 i) {\rm (GeV/c)}^2$
with residues  ${\rm res}[\sigma_s]=(9.05\, 10^{-3} \mp  0.491 i)$ GeV
and ${\rm res}[\sigma_v]=0.261 \mp  0.538 i $ for $\sigma_s$ and $\sigma_v$, respectively.
The second pole, located at $p^2=(-0.507 \pm 3.35 i)~{rm (GeV/c)}^2$ (with the respective residues
${\rm res}[\sigma_s]=(5.5\, 10^{-2} \pm  0.10 i)$ GeV and
  ${\rm res}[\sigma_v]=1.34, 10^{-2} \mp  6.12, 10^{-2}i $), is located already too far from the
corresponding parabola for strange mesons and, consequently is irrelevant in numerical calculations.

With these calculations our analysis of the pole structure is completed. Let us remind the
prepositions: (i) tDS equation Eq.~(\ref{sde}), restricted to the momentum range relevant for mesons as $q\bar q$
bound states, (ii)  tBS equations and $M_{q\bar q}< 3$  GeV, (iii) combined vertex-gluon kernel (\ref{phenvf}) with
IR part only.

\section{Impact of the UV term}
\label{UVimpact}
 An unpleasant fact    is to be noted here. As seen from the Table~\ref{tb2} the fourth pole  is
  located exactly on the real axis,
 hence formally violating   confinement. Presumably
 this is   due to the  simplified interaction we used so far, i.e.  ignoring the UV term in Eq.~(\ref{phenvf}).
 We reiterate  that, in spite of the minor contribution of the UV term for meson masses with light quarks
 according to Refs.~\cite{souglasInfrared,Alkofer,ourFB},
 it can formally improve the behaviour of the solution on the real axis. Indeed,
 an analysis of the solution of the tDS equation  along the real $p^2$ axis
   with the UV term taken into account shows that, at
least in the considered   domain, the pole singularities of the propagator functions disappear.
  However, instead it turns out that with both, IR and  UV, terms the new solutions
  $A(p)$ and $B(p)$ are not anymore  analytical  even for Re\,$p^2 > 0$, as evidenced by
    logarithmic branch point singularities mentioned in~\cite{intJMP}. In the present paper we
  do not discuss the physical meaning and mathematical implications of such singularities in the complex plane.
  Instead,  we try to  reconcile the absence of singularities on the real axis with analyticity
 of $A(p)$ and $B(p)$ in the complex plane, i.e. to keep the logarithmic term only either on
 or in a close vicinity of the real $p^2$ axis making it vanishing at larger Im\,$p^2$.
 This can be accomplished by multiplying the UV term
 by a  damping function which is equal to unity on the real axis and vanishes elsewhere. Such a procedure
 has been proposed, e.g. in Ref.~\cite{PenningtonUV}, to manipulate the IR and UV structure of the ghosts in
 dressing the gluon propagator.  In such a manner the solution along the real axis
 preserves  the correct ultraviolet behaviour. Then, the new solution
  is used to compute the tDS integrals in~(\ref{a}) and (\ref{ab1}) in the complex plane  with the IR term only.
  Such a modification of the phenomenological truncation of the DS equation looks  somehow
 artificial, however, it maintains  the correct UV behaviour of the solution on the real axis and
 provides, as before,  analytical solutions $A(p)$ and $B(p)$ in the complex plane and does not affect
 the solution of the tBS equation for $M_{q\bar q} \lesssim 2$  GeV.
 Certainly,  the solution and the propagator functions will slightly differ from the ones previously
 obtained with the IR term only. Repeating  all the above analysis for the new solution
 one finds  the new positions and residues for the modified kernel. The result is listed in Table \ref{tb3}.
Note that  including  the UV  term in~(\ref{phenvf}) with additional parameters $\tau$, $\gamma_m$ and  $\lambda_{QCD}$
 requires  slightly modified  parameters $D$ and $\omega$, cf.~Ref.~\cite{Maris:2003vk}.
\begin{table}[!ht]
\caption{The pole structure of  the propagator
functions for   $u$, $d$ quarks ($m_q=5$\, MeV).
The effective parameters for the  kernel (\ref{phenvf})
are from Refs. \cite{Maris:2003vk,Alkofer,Roberts}:
  $\omega=0.4$ GeV, $D=36.45~{\rm GeV^{-2}}$, $\tau=e^2-1$, $\Lambda_{QCD}=0.234~{\rm  GeV}$,
$\gamma_m=0.48$.  Notation as in Table~\ref{tb2}. For completeness, the corresponding poles and residues
 for the $s$ quark are also displayed (bottom part).}
\begin{tabular}{lccccc} \hline\hline
      $u$, $d$ quarks              &             1           &        2               &            3                  \\
  pole position     &  (-0.215,$\pm$ 0.335)        &  (-0.12, $\pm$ 2.21)   &      (-1.091,$\pm$ 0.49)     \\
 res[$\sigma_s$ ]   & (-0.017, $\mp$  0.325)       & (0.035 , $\pm$ 0.09)   &    (-0.021, $\mp$ 0.0017)     \\
 res[$\sigma_v$]    & (0.224,  $\mp$ 0.464)        & (0.02,    $\mp$ 0.057) &     (-0.002,  $\mp$ 0.003)    \\     \hline
\label{tb3}

      $s$ quark         &             1               &        2              &            3                  \\
  pole position     &  (-0.21, $\pm$  0.41)       &  (-0.14,$\pm$ 2.17)   &    (-0.743  ,$\pm$ 2.53)     \\
 res[$\sigma_s$ ]   & (-0.001, $\mp$  0.30)       & (0.039, $\pm$ 0.08)   &    (-0.039, $\mp$  0.06)     \\
 res[$\sigma_v$]    & (0.23,   $\mp$ 0.38)        & (0.02,  $\mp$ 0.058)  &    (0.004 ,$\mp$ 0.044)    \\                     \hline\hline
\end{tabular}
\end{table}
Comparing with Table~\ref{tb2} with the IR term solely
it is seen that the positions of the poles and the residues of the
propagator functions are a bit different. However, since the
  new solution of the   DS equation, $\sigma_{s,v}(p^2)$, is correlated with
its analytical part and pole structure, cf.~Eq.~(\ref{main}),
this circumstance does not affect the results of solving  the BS equation, implying
for instance the robustness of the numerical results reported in~\cite{ourFB}.

\section{Summary}
\label{summary}
 We analyse analytical properties of the solution of the truncated Dyson-Schwinger (tDS) equation
 for the quark propagator  in the Euclidean
 complex momentum domain which is determined by the truncated Bethe-Salpeter equation
 for  $q\bar q$ bound states with light quarks.
 It is found that, within the ladder rainbow truncation with only the infrared term
 in the combined effective vertex-gluon kernel, the solutions
 $A(p^2)$, $B(p^2)$ and the propagator functions $\sigma_{s,v}(p^2)$ for $c$ quarks
 are analytical functions in the whole considered domain for $M_{q\bar q}< 3$ GeV, while
 for $u$, $d$ and  $s$ quarks they are analytical only
 in the right hemisphere Re\,$p^2 > 0$. At negative Re\,$p^2$, the functions
 $A(p^2)$ and  $B(p^2)$ are still analytical, however, the propagator functions
 $\sigma_{s,v}(p^2)$ contain pole singularities.
 The exact position of the poles and the corresponding residues of the propagator
 functions can be found by applying
 Rouch\'e's  theorem and computing the Cauchy integrals.  Prior to that, in order to localize
 the approximate region with singularities, we analyse the vector fields of the
 inverse propagators, the vortices of which indicate the positions of null vectors.

 The position of the first few poles and the corresponding residues are found with  good accuracy
 relevant to be used in further calculations of the Bethe-Salpeter  (BS) equation. It is also found that, with
 only the effective infrared  term in the parametrization of the combined  vertex-gluon
kernel, the propagator functions exhibit poles on the real axis, formally violating  the
 confinement. These singularities can be removed by taking into account an ultraviolet  term.
 This term  is known to have a minor contribution
 for  not too heavy meson masses and, at the same time, to provide additional
 singularities of  logarithmic branch point types in the  solution of the tDS equation in the complex plane.
 These singularities hamper a simple analysis of the pole structure of the solution. Nevertheless, since the
 ultraviolet term
 is needed mainly to guarantee  the correct asymptotic behaviour along the real axis, we suggest to
 take it into account only when solving the tDS equation for real $p^2$ and then to use such solutions to compute
 the tDS integrals in the complex plane with IR term only. In such a way one can assure the absence of poles  on the real axis and
 analytical functions
 determining the quark propagator  in the considered complex domain.

 The performed analysis is aimed at elaborating  adequate
 numerical algorithms to solve the BS equation in presence of singularities and to investigate the properties of
 mesons, such as the open charm $D$ mesons, related directly to  physical programmes envisaged, e.g. at FAIR.
Furthermore, the knowledge of the analytical structure of the quark propagators is important for
 designing appropriate  phenomenological kernels since it is related to such fundamental characteristics of QCD as
confinement and dynamical chiral symmetry breaking phenomena encoded in the chiral condensate being the trace of the quark
propagator.

\section*{Acknowledgments}
This work was supported in part by the Heisenberg - Landau program
of the JINR - FRG collaboration, GSI-FE and BMBF. LPK thanks for warm hospitality at the
Helmholtz Centre Dresden-Rossendorf. The authors gratefully acknowledge challenging  questions of M. Lutz
on the DS-BS project. We thank M. Viebach for his assistance during the work.

\end{document}